\documentclass[preprintnumbers, floatfix,letterpaper,aps,prd,epsfig,nofootinbib,
longbibliography,
twocolumn
]{revtex4-1}
\usepackage{bm,graphicx,dcolumn,epstopdf,epsf, latexsym,mathbbol, amssymb,amsmath,color,slashed, mathrsfs,mathcomp,simplewick}
\usepackage[center]{subfigure}
\usepackage{multirow}
\usepackage{makecell}
\usepackage{ytableau}
\usepackage{booktabs}
\usepackage[colorlinks,linkcolor=blue,citecolor=blue,urlcolor=blue]{hyperref}

\begin{document}
\allowdisplaybreaks
 \newcommand{\bq}{\begin{equation}}
 \newcommand{\eq}{\end{equation}}
 \newcommand{\bqn}{\begin{eqnarray}}
 \newcommand{\eqn}{\end{eqnarray}}
 \newcommand{\nb}{\nonumber}
 \newcommand{\lb}{\label}
 \newcommand{\f}{\frac}
 \newcommand{\p}{\partial}
\newcommand{\PRL}{Phys. Rev. Lett.}
\newcommand{\PLB}{Phys. Lett. B}
\newcommand{\PRD}{Phys. Rev. D}
\newcommand{\CQG}{Class. Quantum Grav.}
\newcommand{\JCAP}{J. Cosmol. Astropart. Phys.}
\newcommand{\JHEP}{J. High. Energy. Phys.}
\newcommand{\bea}{\begin{eqnarray}}
\newcommand{\ena}{\end{eqnarray}}
\newcommand{\beqa}{\begin{eqnarray}}
\newcommand{\eeqa}{\end{eqnarray}}
\newcommand{\red}{\textcolor{red}}

\newlength\scratchlength
\newcommand\s[2]{
  \settoheight\scratchlength{\mathstrut}%
  \scratchlength=\number\numexpr\number#1-1\relax\scratchlength
  \lower.5\scratchlength\hbox{\scalebox{1}[#1]{$#2$}}%
}


\title{ Gravitational radiations from periodic orbits around Einstein-\AE{}ther black holes}

\author{Shuo Lu${}^{a, b}$}
\email{lushuo@zjut.edu.cn}

\author{Tao Zhu${}^{a, b}$}
\email{Corresponding author: zhut05@zjut.edu.cn}

\affiliation{${}^{a}$Institute for Theoretical Physics \& Cosmology, Zhejiang University of Technology, Hangzhou, 310023, China\\
${}^{b}$ United Center for Gravitational Wave Physics (UCGWP),  Zhejiang University of Technology, Hangzhou, 310023, China}

\date{\today}

\begin{abstract}

In this work, we investigate the gravitational wave emission from the periodic orbital motion of a test particle around two specific types of black holes in Einstein-\AE{}ther theory, a modified gravity that locally breaks Lorentz symmetry while remaining consistent with theoretical and observational constraints through a careful selection of its four coupling constants $c_i$. Focusing on the impact of the \ae{}ther field, we examine the properties of periodic orbits, which are characterized by a set of three topological integers $(z, w, v)$ that uniquely classify their trajectories. We then calculate the gravitational waveforms generated by these periodic orbits, identifying potential observational signatures. Our analysis reveals a direct connection between the zoom-whirl orbital behavior of the small compact object and the gravitational waveforms it emits: higher zoom numbers lead to increasingly intricate waveform substructures. Moreover, the presence of the \ae{}ther field introduces significant modifications to these waveforms, imprinting measurable deviations that could be potentially tested or constrained by future space-based gravitational wave detectors. 

\end{abstract}


\maketitle

\section{Introduction}
\renewcommand{\theequation}{1.\arabic{equation}} \setcounter{equation}{0}

The advent of gravitational wave astronomy, marked by the groundbreaking detection of gravitational waves by LIGO and Virgo in 2016, has opened a new frontier in our exploration of the universe \cite{LIGOScientific:2016aoc, LIGOScientific:2016vbw, LIGOScientific:2016vlm, LIGOScientific:2016emj}. These spacetime ripples, predicted by Einstein's general theory of relativity, offer a unique observational window into the most energetic and violent cosmic events, such as binary black holes and binary neutron star mergers. Beyond these cataclysmic phenomena, the study of particle trajectories around black holes provides a powerful theoretical framework for probing the intricate dynamics of strong gravitational fields. Among these trajectories, periodic orbits are of particular significance because of their role in addressing fundamental challenges in astrodynamics. The analysis of periodic orbits not only sheds light on the stability of celestial systems and the complex interactions between black holes and their surrounding matter but also captures fundamental information for understanding generic orbital dynamics \cite{Levin:2008mq, Levin:2009sk, Misra:2010pu, Babar:2017gsg}. All generic orbits around black holes can be considered as small deviations from periodic orbits \cite{Levin:2008mq}. The study of periodic orbits and their gravitational wave emissions is of particular interest as well because of their potential observational applications in future space-based gravitational wave detectors. 

Black holes with stellar mass or neutron stars are often found in close orbits around supermassive black holes. Such binary systems are known as the extreme mass ratio inspiral (EMRI), being one of the most important targets of future space-based gravitational detectors, such as Taiji \cite{Hu:2017mde}, Tianqin \cite{TianQin:2015yph, Gong:2021gvw}, and LISA \cite{Danzmann:1997hm, Schutz:1999xj, Gair:2004iv, LISA:2017pwj, Maselli:2021men}, etc. Given that the energy carried away by the lower-mass object’s orbital motion is an exceedingly small fraction of the system’s total energy, the time it takes for the smaller-mass object to spiral around the supermassive black hole can span several years. During this process, the orbital dynamics of the smaller-mass object can be well-approximated by the periodic orbits.  

To characterize the periodic orbits, there is a classification of periodic orbits for massive particles, which is highly useful for understanding the dynamics of black hole mergers \cite{Levin:2008mq}. The main idea of this classification scheme is that a dynamic system could be understood by studying its periodic orbits. To be exact, there are three topological integers indexing all closed orbits around a black hole, representing scaling ($z$), rotation ($\omega$), and vertex ($\nu$) behaviors, respectively. Under this taxonomy, extensive research has been carried out on periodic orbits within various black hole spacetimes, to mention a few, including those of Schwarzschild and Kerr \cite{Levin:2008ci, Levin:2009sk, Bambhaniya:2020zno, Rana:2019bsn}, charged black hole \cite{Misra:2010pu}, naked singularities \cite{Babar:2017gsg}, Kerr-Sen black holes \cite{Liu:2018vea}, and hairy black holes in Horndeski' s theory \cite{Lin:2023rmo}. For the studies of periodic orbits in other black holes, see refs.~\cite{Yao:2023ziq, Lin:2022llz, Chan:2025ocy, Wang:2022tfo, Lin:2023eyd, Haroon:2025rzx, Habibina:2022ztd, Zhang:2022psr, Lin:2022wda, Gao:2021arw, Lin:2021noq, Deng:2020yfm, Tu:2023xab, Zhou:2020zys, Gao:2020wjz, Deng:2020hxw, Azreg-Ainou:2020bfl, Wei:2019zdf, Pugliese:2013xfa,Zhang:2022zox, Healy:2009zm, Wang:2025wob}.
The gravitational wave emissions from the periodic orbits of a large number of black hole spacetimes have also been studied; see refs. \cite{Tu:2023xab, Yang:2024lmj, Shabbir:2025kqh, Junior:2024tmi, Zhao:2024exh, Jiang:2024cpe, Yang:2024cnd, Meng:2024cnq, Li:2024tld, QiQi:2024dwc, Haroon:2025rzx, Alloqulov:2025ucf, Wang:2025hla} and references therein. 

In this paper, we investigate the gravitational wave emission from the periodic orbital motion of a test particle around a black hole in Einstein-\AE{}ther theory. While the periodic orbits around the Einstein-\AE{}ther black hole have been previously studied in \cite{Azreg-Ainou:2020bfl}, here our purpose is to explore the periodic orbits in more detail and focus on the corresponding gravitational wave radiations. Einstein-\AE{}ther theory is a generally covariant theory of gravity which breaks the Lorentz symmetry locally. This theory introduces a timelike vector field, the \ae{}ther field, which provides a preferred timelike direction but remains consistent with current theoretical and observational constraints by properly choosing its four coupling constants $c_i$'s \cite{Jacobson:2007veq, Eling:2004dk}. Perturbative analysis of Einstein–\AE{}ther theory on a Minkowski background, with all coupling constants $c_1,c_2,c_3,c_4$ non-zero, indicates the existence of one scalar, two vectors, and two tensor propagating degrees of freedom \cite{Jacobson:2004ts}.

The astrophysical implications of the Einstein-\AE{}ther theory have been extensively explored in the literature, see refs. \cite{Eling:2006df,Rayimbaev:2022qlx,Streibert:2024cuf,Mukherjee:2024hht,Zhang:2019iim,Liu:2021yev,Zhang:2023kzs,Azreg-Ainou:2020bfl,Hou:2023pfz,Gupta:2021vdj,Foster:2005dk,DeFelice:2024dbj,Berglund:2012bu,Mukohyama:2024vsn,Foster:2007gr,Taherasghari:2025mlf,Foster:2006az,Zhu:2019ura} and references therein. The observational constraints on its four coupling constants are also investigated with the data of the gravitational wave events GW170817 and GRB 170817A \cite{Oost:2018tcv, Gong:2018cgj}, which put a stringent limit $|c_T-1|\lesssim10^{-15}$ on the tensor propagation speed $c_T$ \cite{Arcavi:2017xiz}, thereby translating to the bound $|c_1+c_3|\lesssim10^{-15}$ \cite{Oost:2018tcv}. 

The main purpose of this article is to study the periodic orbital behaviors of a particle surrounding a black hole in Einstein-\AE{}ther theory and their corresponding gravitational wave radiations. We explore how \AE{}ther effect affects the behaviors of orbits and calculate the corresponding gravitational wave radiation. 
The article is constructed as follows. In the section.\ref{section2}, we present a brief review of Einstein-\AE{}ther theory and two specific spherical symmetric black hole solutions, and then in Sec.~\ref{section3} we discuss the geodesics of a massive test particle around the black holes in Einstein-\AE{}ther theory and study the corresponding periodic orbits. In the section.\ref{section4}, we calculate the gravitational wave radiation of periodic orbits around the black holes in the Einstein-\AE{}ther theory. The conclusion and discussion are presented in Sec .~\ref {section5}.

\section{Black holes in Einstein-\AE{}ther theory}\label{section2}
\renewcommand{\theequation}{2.\arabic{equation}} \setcounter{equation}{0}

In this section, we present a brief review of the black hole solutions in the Einstein-\AE{}ther theory.

\subsection{Field Equations of the Einstein-\AE{}ther theory}

Einstein-\AE{}ther theory is a Lorentz-violating modification of general relativity that includes a dynamical timelike unit vector field $u^\mu $, known as the \ae{}ther field, which defines a preferred local rest frame \cite{Jacobson:2007veq}. The theory retains diffeomorphism invariance while explicitly breaking local Lorentz symmetry. The total action of the theory is described by the action \cite{Jacobson:2007veq},
\bqn\label{eq:action}
S_{\ae}=\frac{1}{16\pi G_{\ae}}\int d^4x\sqrt{-g}(R+\mathcal{L}_{\ae}) + \int d^4x \sqrt{-g} {\cal L}_{\rm M}, \nb\\
\eqn
where $g$ is the determinant of the four-dimensional metric $ g_{\mu\nu} $ with the signatures $(-,+,+,+)$, $R$ is the Ricci scalar, $G_{\ae}$ is the \ae{}ther gravitational constant, ${\cal L}_{\rm M}$ is the Lagrangian of the matter field, and the Lagrangian of the \ae{}ther field $\mathcal{L}_{\ae}$ is defined as 
\bqn
\mathcal{L}_{æ}\equiv -M^{\alpha \beta }_{{\mu \nu }} (D_{\alpha} u^{\mu })(D_{\beta }u^{\nu})+\lambda(g_{\alpha\beta}u^{\mu}u^{\nu}+1),
\eqn
Here, $ D_\alpha $ is the covariant derivative, and the coefficients $ M^{\alpha\beta}_{\phantom{\alpha\beta}\mu u} $ are defined in terms of four coupling constants $ c_i $ as:
\begin{equation}
M^{\alpha\beta}_{\phantom{\alpha\beta}\mu
\nu} \equiv c_1 g^{\alpha\beta} g_{\mu
\nu} + c_2 \delta^\alpha_\mu \delta^\beta_
\nu + c_3 \delta^\alpha_
\nu \delta^\beta_\mu - c_4 u^\alpha u^\beta g_{\mu
\nu}.
\end{equation}
The Lagrange multiplier $\lambda $ enforces the normalization condition $ u^\mu u_\mu = -1$. The source-free Maxwell Lagrangian $\mathcal{L}_M$ is given by
\bqn
\mathcal{L}_M&=&-\frac{1}{16\pi G_{æ}}\mathcal{F}_{\alpha\beta}\mathcal{F}^{\alpha\beta},\\
\mathcal{F}_{\alpha\beta}&=&D_\alpha \mathcal{A}_\beta-D_\beta\mathcal{A}_\alpha,
\eqn
where $\mathcal{A}_a$ is the electromagnetic potential four-vector.

Varying the action \eqref{eq:action} with respect to the metric $ g_{\mu
\nu} $, the \ae{}ther field $ u^\mu $, and the Lagrange multiplier $\lambda$ lead to the field equations:
\bqn
&&R_{\mu \nu} - \frac{1}{2} g_{\mu\nu} R= 8 \pi G_{\ae} (T_{\mu\nu}^{\ae} + T_{\mu\nu}^{\rm M}), \\
&&D_\mu J_\alpha^\mu+c_4 a_\mu D_\alpha u^\mu+\lambda u_\alpha=0, \lb{aether_eq}\\
&& g_{\mu \nu} u^\mu u^\nu=-1, \lb{lambda_eq}
\eqn
where
\bqn
T_{\alpha \beta}^{\ae} &\equiv & D_\mu\left[J_{(\alpha}^\mu u_{\beta)}+J_{(\alpha \beta)} u^\mu-u_{(\beta} J_{\alpha)}{ }^\mu\right] \nb\\
&& +c_1\left[\left(D_\alpha u_\mu\right)\left(D_\beta u^\mu\right)-\left(D_\mu u_\alpha\right)\left(D^\mu u_\beta\right)\right] \nb\\
&& +c_4 a_\alpha a_\beta+\lambda u_\alpha u_\beta-\frac{1}{2} g_{\alpha \beta} J_\sigma^\delta D_\delta u^\sigma, \\
T_{\alpha \beta}^{M}&=&\frac{1}{4\pi G_{æ}}\left[-\frac{1}{4}g_{\alpha\beta} \mathcal{F}_{mn}\mathcal{F}^{mn}+\mathcal{F}_{\alpha m}\mathcal{F}_\beta^m\right],\\
J^\alpha{ }_\mu &\equiv & M^{\alpha \beta}{ }_{\mu \nu} D_\beta u^\nu, \\
a^\mu &\equiv & u^\alpha D_\alpha u^\mu,
\eqn
From Eqs.~(\ref{aether_eq}) and (\ref{lambda_eq}), one finds
\bqn
\lambda=u_\beta D_\alpha J^{\alpha \beta}+c_4 a^2,
\eqn
where $a^2 \equiv a_\lambda a^\lambda$.

The gravitational constant $G_{æ}$ is related to the Newtonian's gravitational constant $G_{N}$ via the relation \cite{Carroll:2004ai}
\bqn
G_{N}=\frac{G_{æ}}{1-\frac{1}{2}c_{14}}, \nb\\
\eqn
where $c_{ij}\equiv c_{i}+c_{j}$. Hereafter, we shall adopt the unit $G_N=1=c$.

\subsection{Static and spherically symmetric Einstein-\AE{}ther Black Holes}

The general form of a static spherically symmetric metric for the Einstein-\AE{}ther black hole in the Eddington-Finkelstein coordinates is given by
\bqn\lb{metric_EF}
ds^{2}=-e(r)dv^{2}+2f(r)dvdr+r^{2}d\Omega^2,
\eqn
where $e(r)$ and $f(r)$ are two metric functions, and $d\Omega^2=d\theta^2+\sin^2\theta d\phi^2$. The \ae{}ther field $u^\mu$ can be written as
\bqn
u^{\mu}(r) = \{\alpha(r), \beta(r), 0, 0\}.
\eqn
In the above expressions, $e(r)$, $f(r)$, $\alpha(r)$, and $\beta(r)$ are functions of $r$ only.

There are two types of exact static and charged spherically symmetric black hole solutions in Einstein-\AE{}ther theory~\cite{Ding:2015kba}, which correspond to two specific combinations of the coupling constants of the \ae{}ther field, i.e., $c_{14}=0$ but $c_{123} \neq 0$ for the first type and $c_{123}=0$  for the second type, respectively. In the following, we present their explicit metric functions, respectively.

For the first solution, which corresponds to $c_{14}=0$ but $c_{123} \neq 0$, its metric functions are given by \cite{Ding:2015kba}
\bqn\lb{e(r)1}
e(r) &=& 1 - \frac{2M}{r}
- \frac{27c_{13}}{256(1-c_{13})}\left(\frac{2M}{r}\right)^{4},  \\
f(r)& =& 1,\\
\alpha(r)&=&\left[\frac{1}{\sqrt{1-c_{13}}} \frac{3\sqrt{3}}{16}\left(\frac{2M}{r}\right)^{2} \right.\nb\\
&&\left.+ \sqrt{1 - \frac{2M}{r} + \frac{27}{256}\left(\frac{2M}{r}\right)^{4}}\right]^{-1},\\
\beta(r) &=& -\frac{1}{\sqrt{1-c_{13}}} \frac{3\sqrt{3}}{16}\left(\frac{2M}{r}\right)^{2}.
\eqn
Here $M$ is the mass of the black hole spacetime. It is easy to see that when $c_{13}=0$, the above solution reduces to the Schwarzschild black hole.

For the second solution, which corresponds to $c_{123}=0$, we have \cite{Ding:2015kba}
\bqn\lb{e(r)123}
e(r)&=&1-\frac{2M}{r} 
-\frac{2c_{13}-c_{14}}{8(1-c_{13})}(\frac{2M}{r})^{2},\\
f(r)&=&1,\\
\alpha(r)&=&\frac{1}{1+\frac{1}{2}\left(\sqrt{\frac{2-c_{14}}{2(1-c_{13})}-1}\right)\frac{2M}{r}},\\
\beta(r)&=&-\frac{1}{2}\sqrt{\frac{2-c_{14}}{2(1-c_{13})}}\frac{2M}{r}.
\eqn
In this case, when $c_{13}=c_{14}=0$, it also reduces to the Schwarzschild black hole.

It is also convenient to transform the Eddington-Finkelstein coordinate system into the form of the usual coordinates $(t, r, \theta, \phi)$. The coordinate transformation read
\bqn
dt = dv -\frac{dr}{e(r)}, \;\; dr=dr.
\eqn
Then the metric (\ref{metric_EF}) turns into the form
\bqn
ds^2= -e(r)dt^2 + \frac{dr^2}{e(r)} +r^2 (d\theta^2 + \sin^2\theta d\phi^2).
\eqn
In this metric, the \ae{}ther field reads
\bqn
u^\mu=\left(\alpha(r)- \frac{\beta(r)}{e(r)}, \beta(r), 0, 0\right).
\eqn

\section{Periodic orbits}\label{section3}
In this section, we discuss the periodic timelike orbits around the Einstein-\AE{}ther black holes. 

Let us first consider the evolution of a particle in the black hole spacetime. We start with the Lagrangian of the particle,
\bqn
{\cal L }= \frac{1}{2}g_{\mu \nu} \frac{d x^\mu} {d \tau } \frac{d x^\nu}{d \tau},
\eqn
where $\tau$ denotes the proper time, which equals the affine parameter of the world line of the timelike particle. For a massless particle, we have ${\cal L}=0$ and for a massive one ${\cal L} <0$. Then the generalized momentum $p_\mu$ of the particle can be obtained via
\bqn
p_{\mu} = \frac{\partial {\cal L}}{\partial \dot x^{\mu}} = g_{\mu\nu} \dot x^\nu,
\eqn
which leads to four equations of motion for a particle with energy $E$ and angular momentum $L_z$,
\bqn
p_t &=& g_{tt} \dot t  = - E,\\
p_\phi &=& g_{\phi \phi} \dot \phi = L_z, \\
p_r &=& g_{rr} \dot r,\\
p_\theta &=& g_{\theta \theta} \dot \theta.
\eqn
Here, a dot denotes the derivative with respect to the affine parameter $\lambda$ of the geodesics. From these expressions, we obtain
\bqn\lb{dot1}
\dot t = - \frac{ E  }{ g_{tt} } = \frac{E}{e(r)},\\
\lb{dot2}\dot \phi = \frac{ L_z}{g_{\phi\phi}} = \frac{L_z}{r^2 \sin^2\theta}.
\eqn
For timelike geodesics, we have $ g_{\mu \nu} \dot x^\mu \dot x^\nu = -1$. Substituting $\dot t$ and $\dot \phi$ we can get
\bqn
g_{rr} \dot r^2 + g_{\theta \theta} \dot \theta^2 &=& -1 - g_{tt} \dot t^2  - g_{\phi\phi} \dot \phi^2\nb\\
&=& -1 +\frac{E^2}{e(r)}- \frac{L_z^2}{r^2\sin^2\theta}.
\eqn

We are interested in the evolution of the particle in the equatorial circular orbits. For simplicity, we choose $\theta=\pi/2$ and $\dot \theta=0$. Then the above expression can be simplified into the form
\bqn\lb{rdot}
\dot r ^2 = E^2 - V_{\rm eff}(r),
\eqn
where $V_{\rm eff}(r)$ denotes the effective potential and is given by
\bqn \lb{Veff}
V_{\rm eff}(r)= \left(1+\frac{L_z^2}{r^2}\right)e(r).
\eqn

One immediately observes that $V_{\rm eff}(r) \to 1$  as $r \to +\infty$, as expected for an asymptotically flat spacetime. In this case, particles with energy $E >1$ are able to escape to infinity, and $E = 1$ is the critical case between bound and unbound orbits. In this sense, the maximum energy for the bound orbits is $E=1$. Thus, we can obtain the trajectory of a particle by integrating Eqs.~(\ref{dot1}), Eqs.~(\ref{dot2}), and Eqs.~(\ref{Veff}) to get $t$, $\phi$, and $r$ as a function of $\tau$. However, 
since Eq.~(\ref{rdot}) involves taking a square root, and the sign must be specified manually before any numerical integration. There is a convenient equation of motion that can be derived for numerical analysis from the $r-$ component of the geodesic equation, which is
\bqn
\ddot{r}=\frac{e'(r)}{2e(r)}\dot{r}^2-\frac{e'(r)E^2}{2e(r)}+\frac{e(r)L_z^2}{r^3}.
\eqn

Upon performing the integration, a periodic orbit can be generated with a specific value of $E$ and $L_z$. A periodic orbit is a special type of bound orbit that returns precisely to its initial position after a finite amount of time. Periodic orbits can exhibit a wide variety of shapes. For the sake of systematic analysis, it is therefore necessary to adopt a classification scheme.

We adopt the taxonomy introduced in~\cite{Levin:2008mq} for indexing all periodic orbits around the Einstein-\AE{}ther black holes with a triplet of topological integers $(z, w, v)$, which denote the zoom, whirl, and vertex behaviors of the orbits. Normally, periodic orbits are those orbits that can return exactly to their initial conditions after a finite time, which requires that the ratio between the two frequencies of oscillations in  $r$-motion and $\ phi$-motion has to be a rational number. And generic orbits around the black hole can be approximated by nearby periodic orbits since any irrational number can be approximated by a nearby rational number. Therefore, exploring the periodic orbits would be very helpful in understanding the structure of any generic orbits and the corresponding radiation of the gravitational waves. 

According to the taxonomy of Ref.~\cite{Levin:2008mq}, one introduces the ratio $q$ between the two frequencies, $\omega_r$ and $\omega_\phi$ of oscillations in the $r$-motion and $\phi$-motion respectively, in terms of three integers $(z, w, v)$ as
\bqn
q \equiv \frac{\omega_\phi}{\omega_r}-1 = w + \frac{v}{z}.
\eqn
To be specific, the integers $(z, w, v)$ have different geometric interpretations: $z$ is the zoom number, representing the number of larger circles of an orbit. $w$ is the whirl number, showing the number of small loops near the central object. The number $v$ is the vertex number, which specifies whether the particle traverses the vertices of the periodic orbit in a clockwise or counterclockwise sequence. It should be noted that the numbers of $z$ and $v$ are relatively prime in order to avoid degeneracy~\cite{Levin:2008mq}. The parameter $q$ measures the degree of periapsis precession beyond that of a simple elliptical orbit, offering insight into the orbit's topology. In addition, the tracing order of the leaves, referring to the sequence of orbital paths or segments, is also considered in this framework. Together, these numbers provide a systematic means of characterizing the complex dynamics of periodic orbits. Since $\frac{\omega_\phi}{\omega_r}=\Delta \phi/(2\pi)$ with $\Delta \phi \equiv \oint d\phi$ being the equatorial angle during a period in $r$, which must be an integer multiple of $2\pi$. Using the geodesic equations of the Einstein-\AE{}ther black holes, $q$ can be calculated via
\bqn
q &=& \frac{1}{\pi} \int_{r_1}^{r_2} \frac{\dot \phi}{\dot r} dr -1\nb\\
&=& \frac{1}{\pi} \int_{x_+}^{x_-} \frac{L_z dx'}{\sqrt{P(x)}}-1,
\eqn
where $r_1$ and $r_2$ are two turning points and $P(x)$ is the polynomial that depends on $e(r)$.

In order to find out the exact two turning points for the integration and simplify the calculations, as there may be multiple roots for $r$, it is helpful to reformulate the geodesic equations using a new variable. First, we combine Eqs.~(\ref{dot2}) and (\ref{rdot}) and change the radial variable to $r=1/x$ , which gives
\bqn
\frac{d\phi}{dx}=\pm \frac{L_z}{\sqrt{P(x)}},
\eqn
where $P(x)$ is a polynomial depending on $e(r)$. To be specific, $P(x)$ is a sixth-degree polynomial for the first solution and a fourth-degree polynomial for the second solution. 

In the first solution, where $e(r)$ is given by Eqs.~(\ref{e(r)1}) , $P(x)$ gives

\bqn
P(x)&=&E^{2}\nb\\
&-&(L_z^{2}x^{2}+1) \left[1-2Mx-16M^{4}x^{4}(\frac{27c_{13}}{256(1-c_{13})})\right].\nb\\
\eqn

Numerical exploration shows that, within the typical parameter ranges considered in this work, the equation $P(x)=0$ generally possesses four real roots and two complex roots. Here we denote these real roots by $x_+$, $x_-$, $x_1$, and $x_2$. For $c_{13}>0$, the roots are ordered as follows
\bqn\lb{order1}
x_2\leq0\leq x_+ \leq x_- \leq x_1.
\eqn
As $c_{13}\to0$, the root $x_2$ tends to $-\infty$, This is continued into $c_{13}<0$, where $x_2$ becomes positive, and the roots are ordered as
\bqn\lb{order2}
x_+\leq x_-\leq x_1\leq x_2.
\eqn
In the second solution, where $e(r)$ is given by Eqs.~(\ref{e(r)123}), $P(x)$ gives
\bqn
P(x)&=& E^{2}\nb\\
&-&(L_z^2x^2+1)\left[1-2Mx-4M^2x^2(\frac{2c_{13}-c_{14}}{8(1-c_{13})})\right].\nb\\
\eqn

In this case, $p(x)$ is a fourth-order polynomial, and its roots of $P(x)=0$ can be obtained exactly. Following the convention of the first solution mentioned above, we denote the roots by $x_+$, $x_-$, $x_1$, and $x_2$. For $\frac{2c_{13}-c_{14}}{8(1-c_{13})}>0$, the roots are ordered as Eq.~(\ref{order1}). On the other hand, for $\frac{2c_{13}-c_{14}}{8(1-c_{13})}<0$, the roots are ordered as Eq.~(\ref{order2}). With all the measures mentioned above, the calculation of $q$ can be simplified.

\begin{figure*}[htbp]
\centering
{
\begin{minipage}[b]{.3\linewidth}
\centering
\includegraphics[scale=0.3]{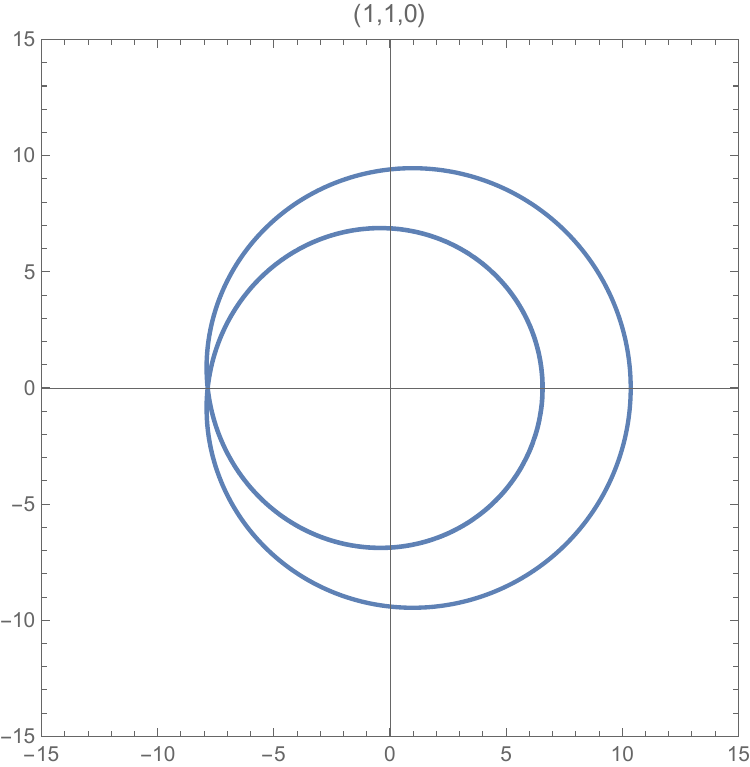}
\end{minipage}
}
{
\begin{minipage}[b]{.3\linewidth}
\centering
\includegraphics[scale=0.3]{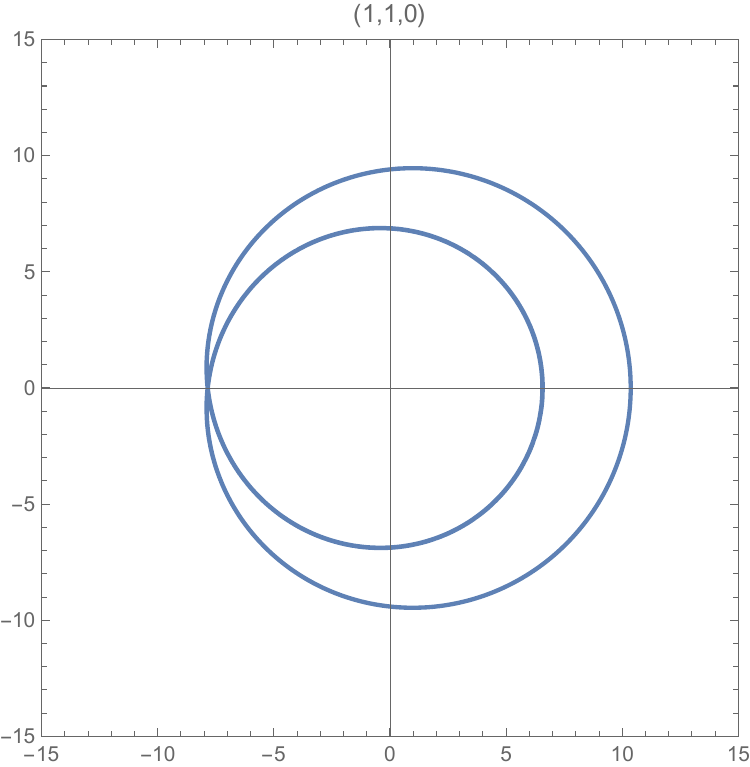}
\end{minipage}
}
{
\begin{minipage}[b]{.3\linewidth}
\centering
\includegraphics[scale=0.3]{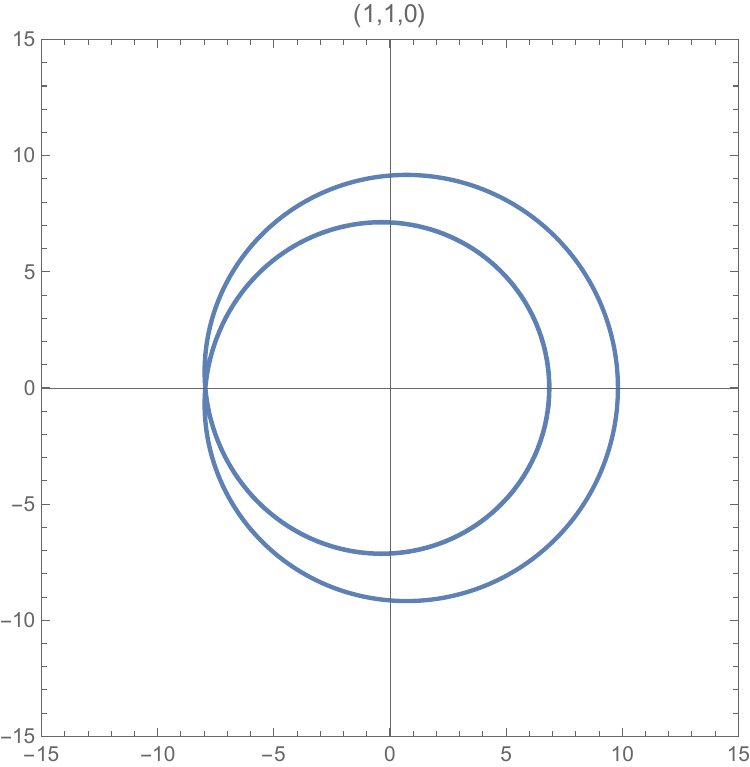}
\end{minipage}
}
{
\begin{minipage}[b]{.3\linewidth}
\centering
\includegraphics[scale=0.3]{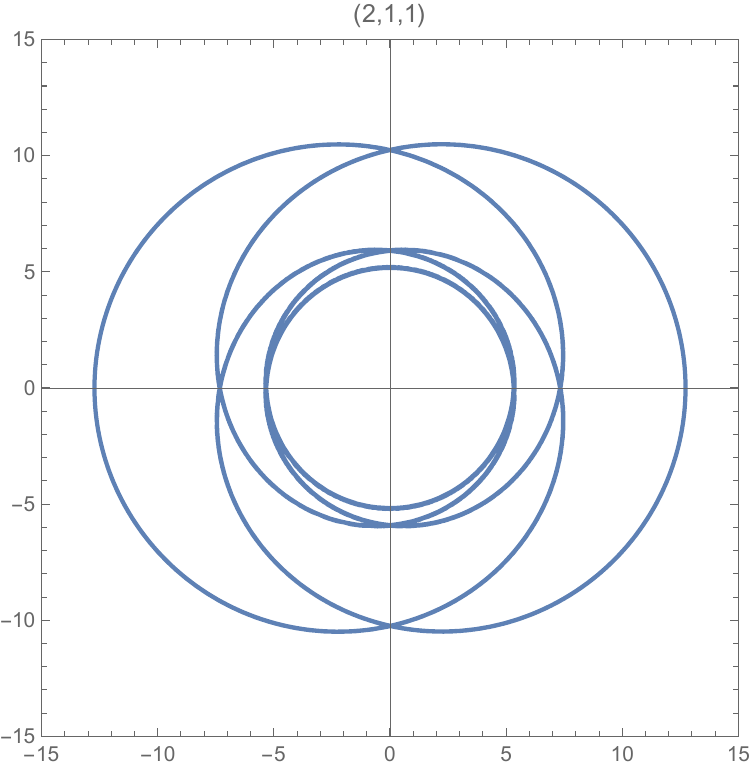}
\end{minipage}
}
{
\begin{minipage}[b]{.3\linewidth}
\centering
\includegraphics[scale=0.3]{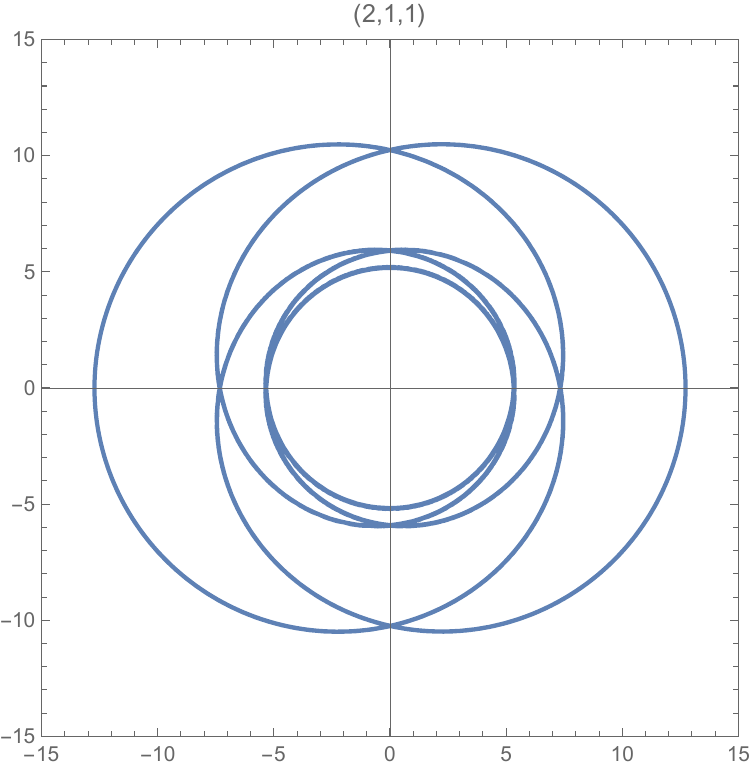}
\end{minipage}
}
{
\begin{minipage}[b]{.3\linewidth}
\centering
\includegraphics[scale=0.3]{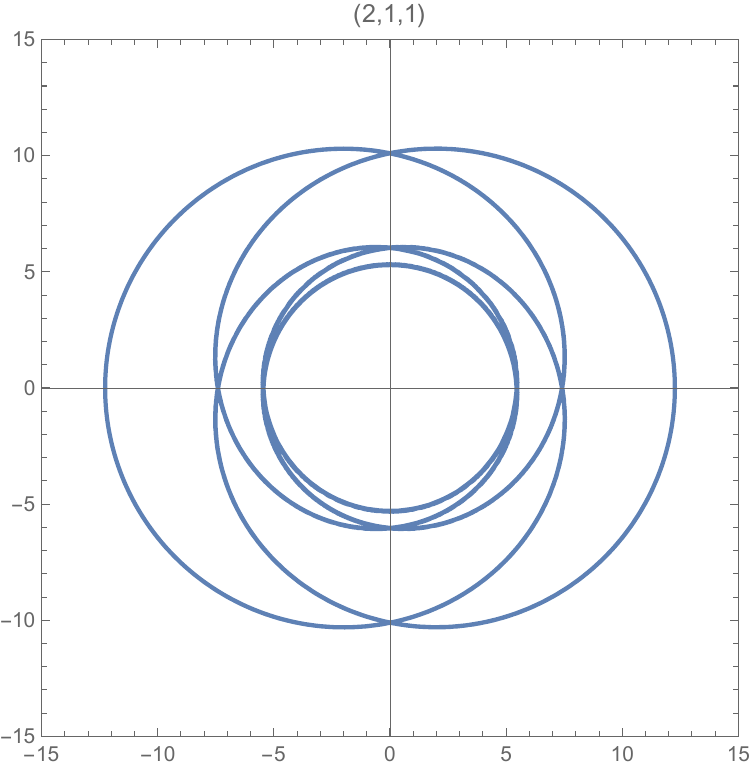}
\end{minipage}
}
{
\begin{minipage}[b]{.3\linewidth}
\centering
\includegraphics[scale=0.3]{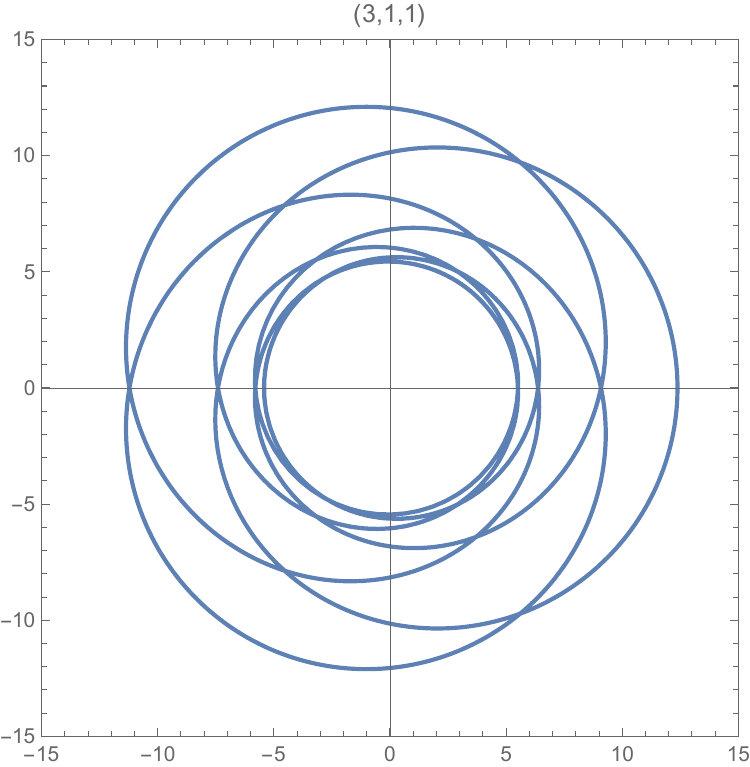}
\end{minipage}
}
{
\begin{minipage}[b]{.3\linewidth}
\centering
\includegraphics[scale=0.3]{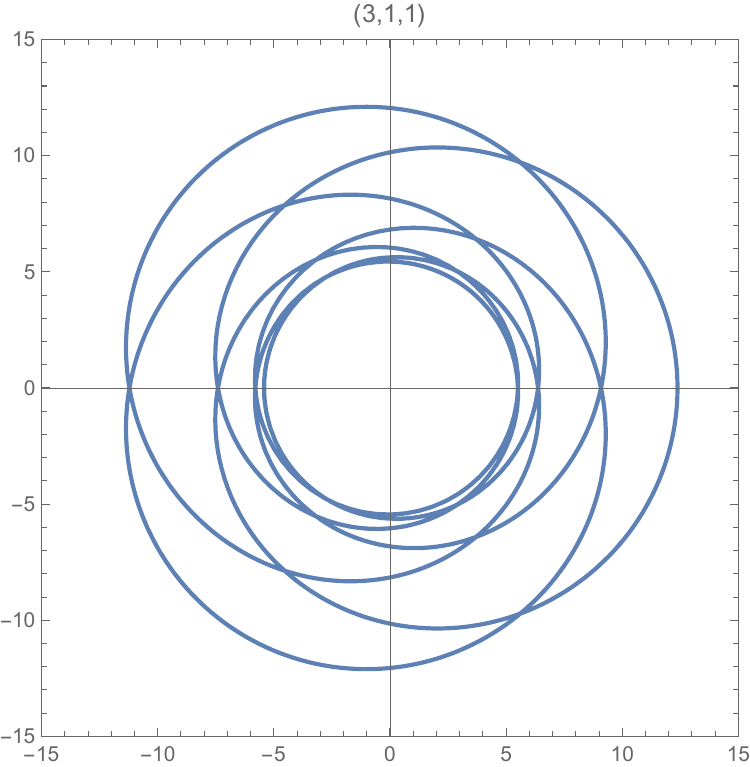}
\end{minipage}
}
{
\begin{minipage}[b]{.3\linewidth}
\centering
\includegraphics[scale=0.3]{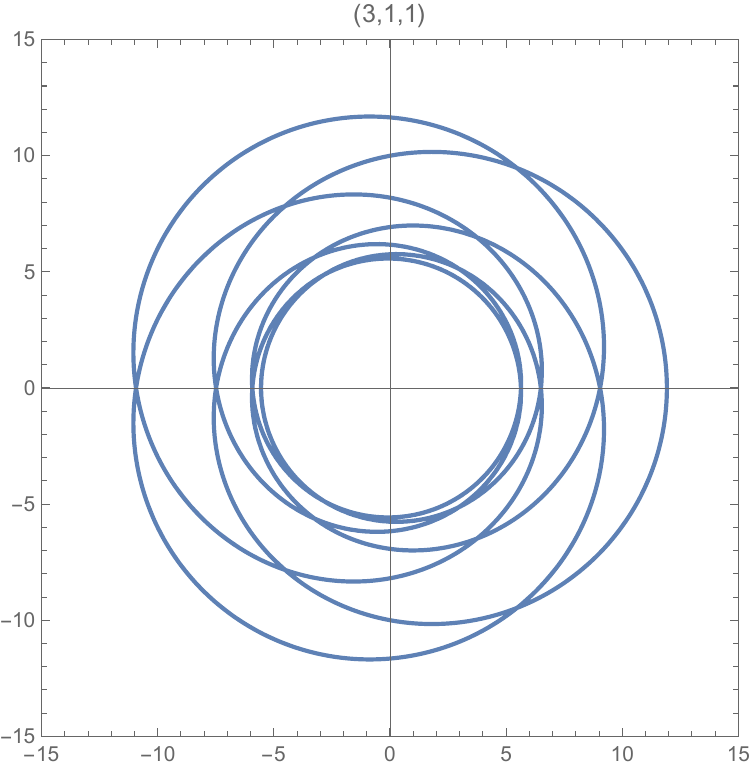}
\end{minipage}
}
{
\begin{minipage}[b]{.3\linewidth}
\centering
\includegraphics[scale=0.3]{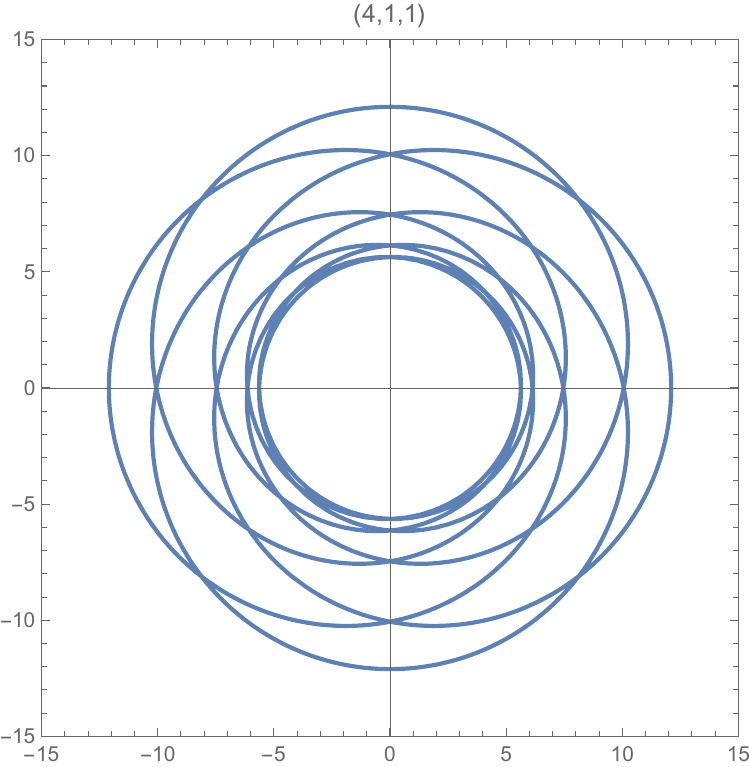}
\end{minipage}
}
{
\begin{minipage}[b]{.3\linewidth}
\centering
\includegraphics[scale=0.3]{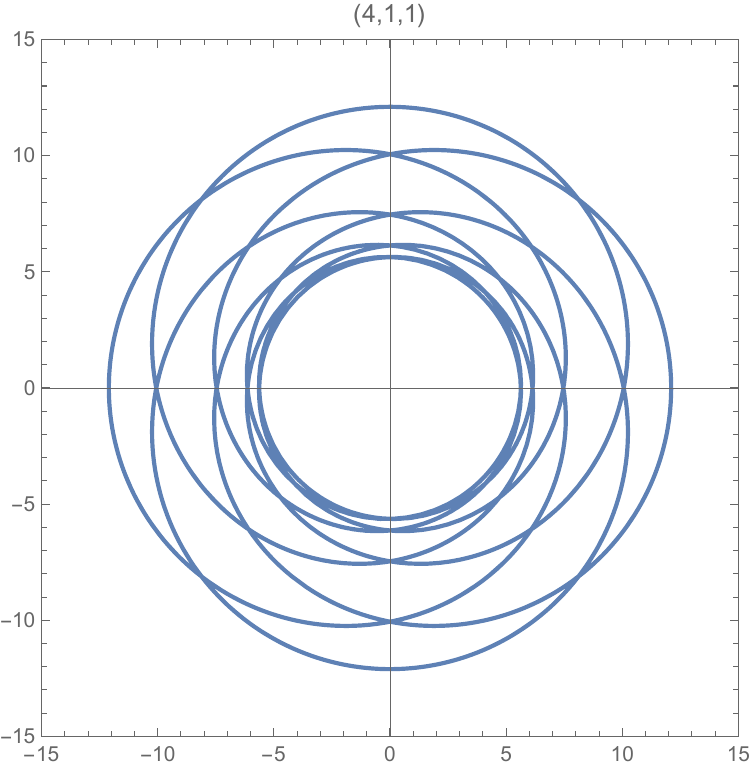}
\end{minipage}
}
{
\begin{minipage}[b]{.3\linewidth}
\centering
\includegraphics[scale=0.3]{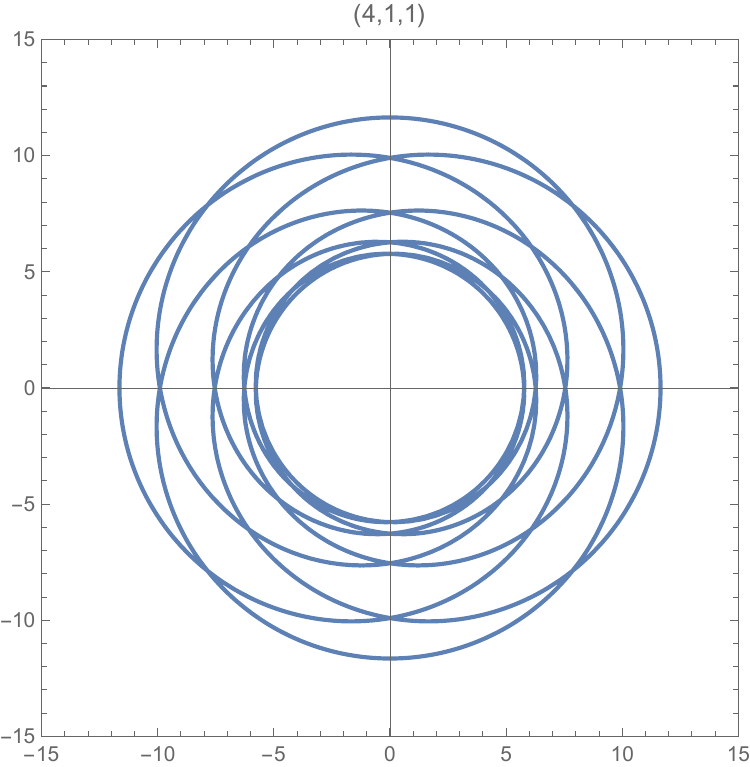}
\end{minipage}
}
\caption{Periodic orbits in first solution of Einstein-\AE{}ther theory. $c_{13}$ are set with different values in each column, which are 0, 0.00002, and 0.2. The values of ($z,w,v$) for each orbit are plotted in the pictures. In this figure, the horizontal and vertical axes are Cartesian coordinates $(x, y) =(r\sin\phi, r \cos\phi)$ with $\theta=\pi/2$.}
\label{firsttype}
\end{figure*}

\begin{figure*}[htbp]
\centering
{
\begin{minipage}[b]{.3\linewidth}
\centering
\includegraphics[scale=0.3]{swz110.pdf}
\end{minipage}
}
{
\begin{minipage}[b]{.3\linewidth}
\centering
\includegraphics[scale=0.3]{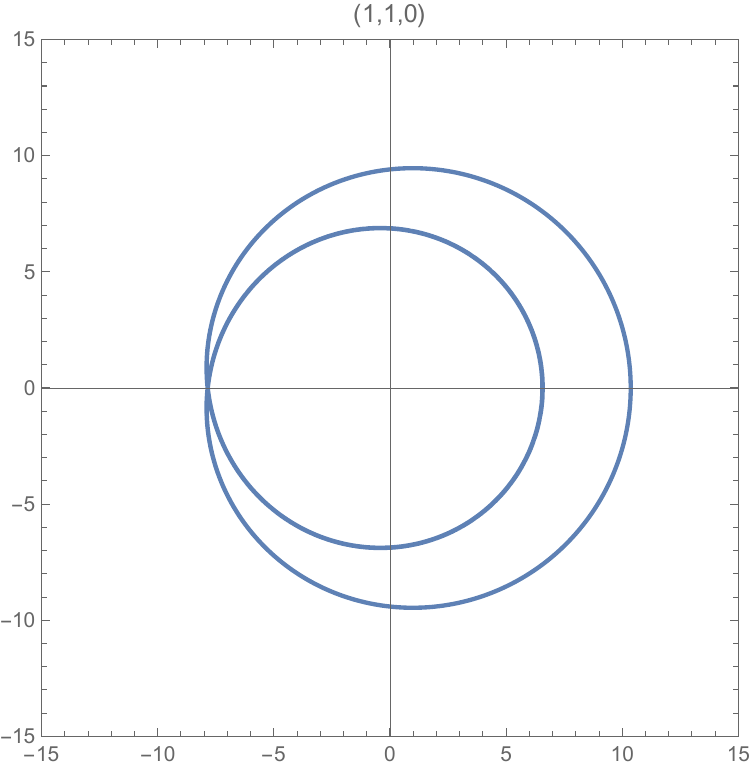}
\end{minipage}
}
{
\begin{minipage}[b]{.3\linewidth}
\centering
\includegraphics[scale=0.3]{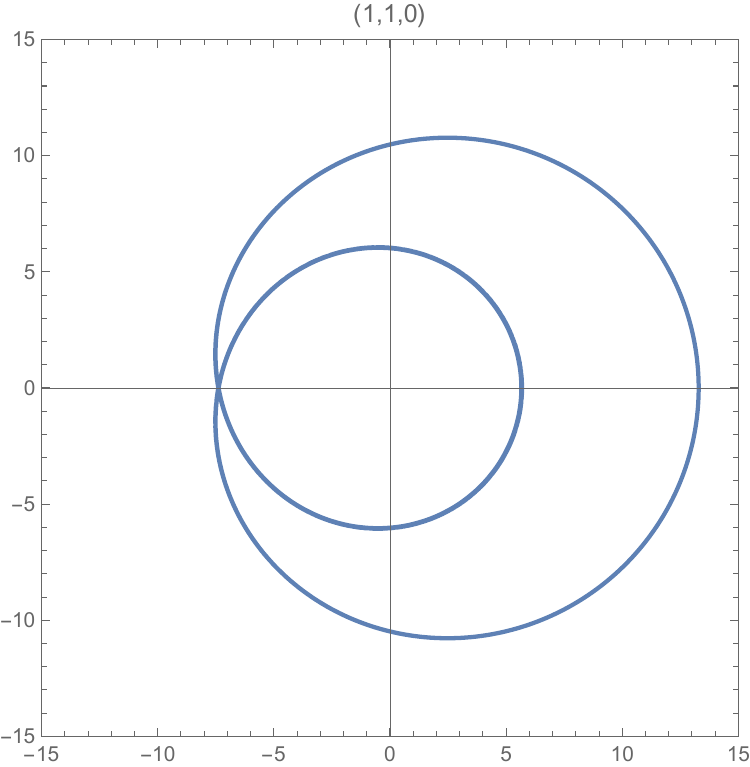}
\end{minipage}
}
{
\begin{minipage}[b]{.3\linewidth}
\centering
\includegraphics[scale=0.3]{swz211.pdf}
\end{minipage}
}
{
\begin{minipage}[b]{.3\linewidth}
\centering
\includegraphics[scale=0.3]{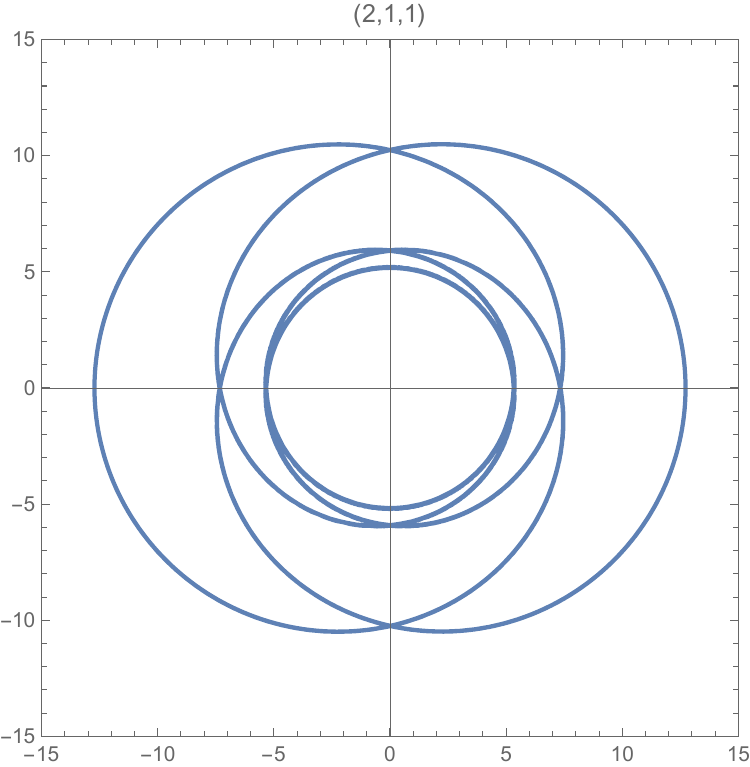}
\end{minipage}
}
{
\begin{minipage}[b]{.3\linewidth}
\centering
\includegraphics[scale=0.3]{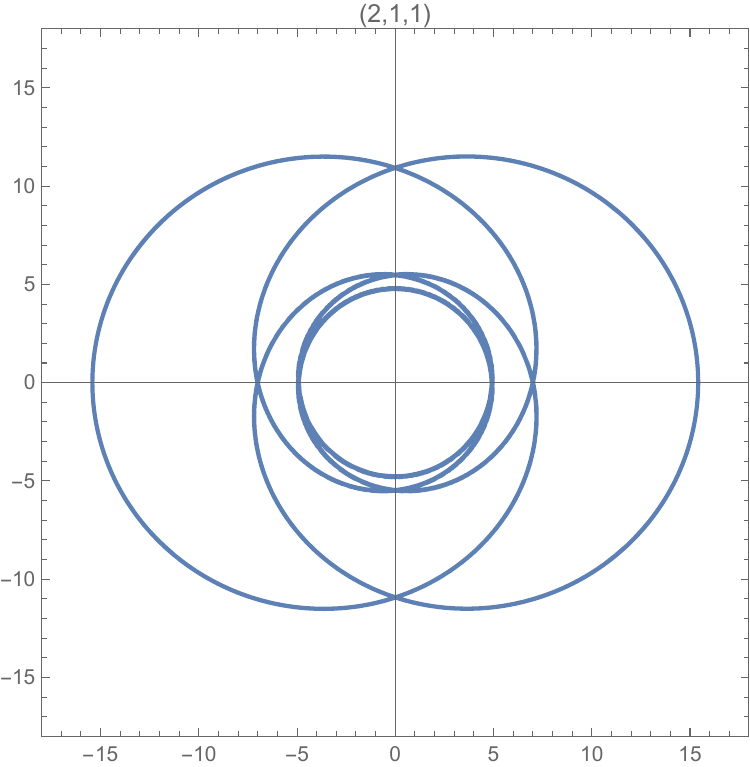}
\end{minipage}
}
{
\begin{minipage}[b]{.3\linewidth}
\centering
\includegraphics[scale=0.3]{swz311.pdf}
\end{minipage}
}
{
\begin{minipage}[b]{.3\linewidth}
\centering
\includegraphics[scale=0.3]{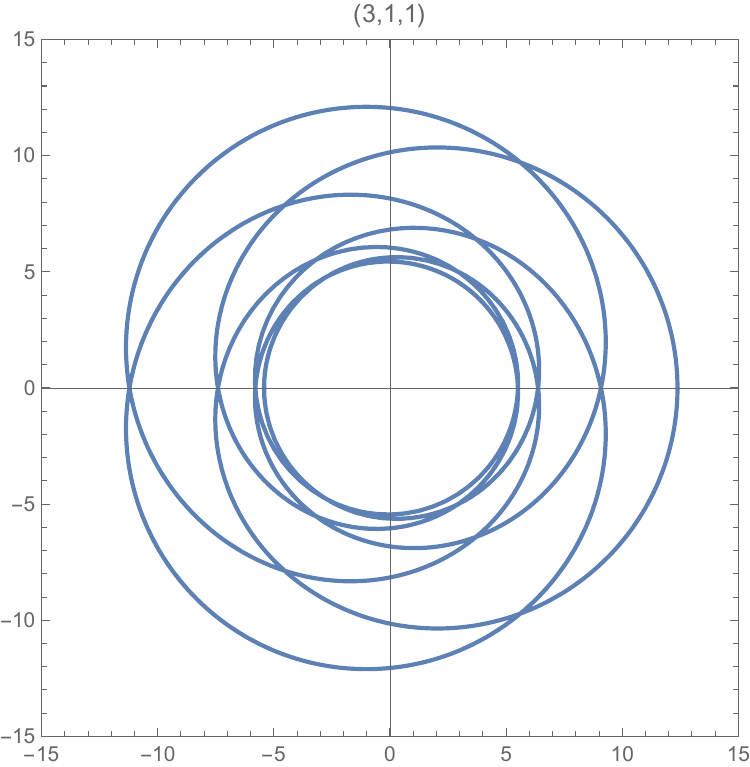}
\end{minipage}
}
{
\begin{minipage}[b]{.3\linewidth}
\centering
\includegraphics[scale=0.3]{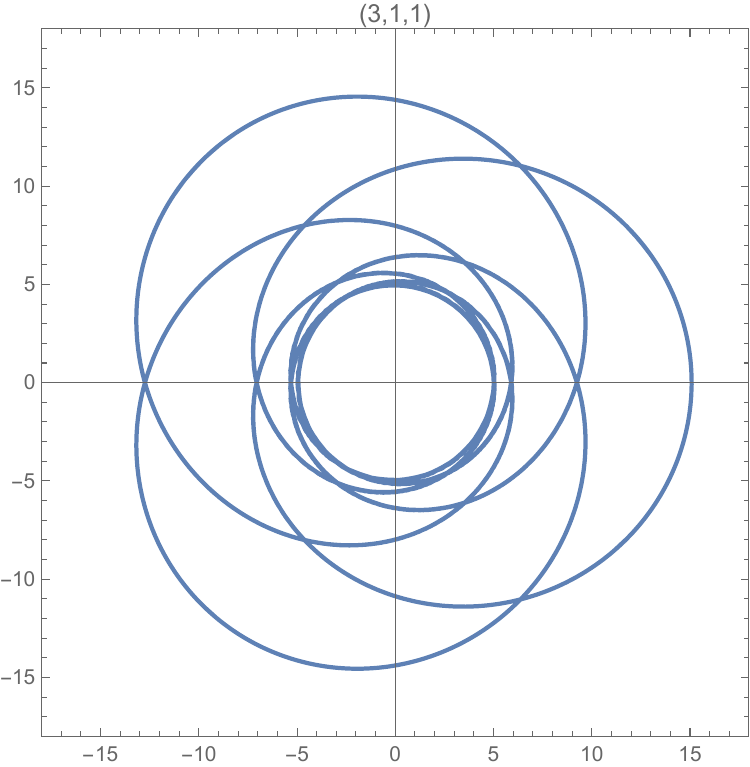}
\end{minipage}
}
{
\begin{minipage}[b]{.3\linewidth}
\centering
\includegraphics[scale=0.3]{swz411.pdf}
\end{minipage}
}
{
\begin{minipage}[b]{.3\linewidth}
\centering
\includegraphics[scale=0.3]{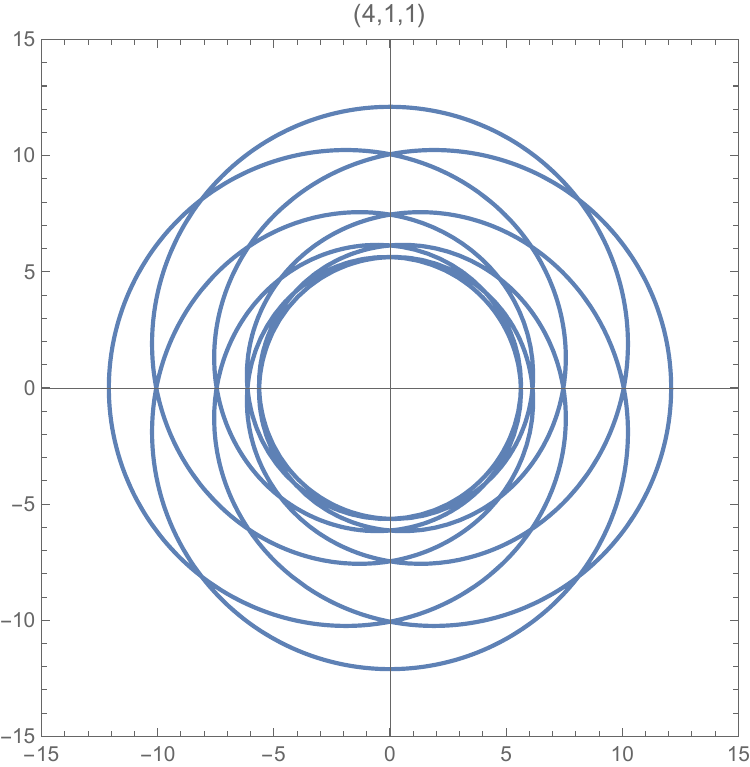}
\end{minipage}
}
{
\begin{minipage}[b]{.3\linewidth}
\centering
\includegraphics[scale=0.3]{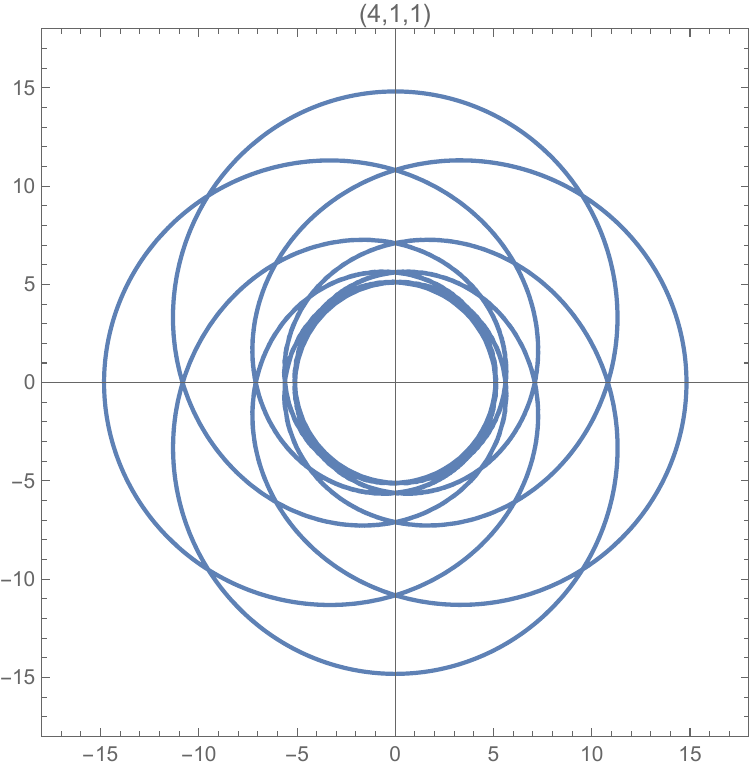}
\end{minipage}
}
\caption{Periodic orbits in second solution of Einstein-\AE{}ther theory. $c_{14}$ are set with different values in each column, which are 0, 0.00002, and 0.2. The values of ($z,w,v$) for each orbit are plotted in the pictures. In this figure, the horizontal and vertical axes are Cartesian coordinates $(x, y) =(r\sin\phi, r \cos\phi)$ with $\theta=\pi/2$.}
\label{secondtype}
\end{figure*}

The behavior of $q$ that varies as $E$ and $L_z$ change can be found in~\cite{Azreg-Ainou:2020bfl}. In Figs.~\ref{firsttype} and \ref{secondtype}, we illustrate the periodic orbits using different combinations of integers $(z, w, v)$ for the two types of Einstein-\AE{}ther black holes, respectively. It is worth mentioning that we set $M=1$ for simplicity in the figures. The value of $z$ determines the number of blades in the orbit’s shape. The larger $z$ values correspond to larger blade profiles and increasingly complex trajectories. Fig.~\ref{firsttype} displays periodic orbits for the first solution (with $c_{14}=0$), and in each column of the figure, we use the same combinations of integers $(z, w, v)$ but with different values of $c_{13}$. The value of $c_{13}$ is set to 0, 0.00002, 0.2 from left to right in each column. In Fig.~\ref{secondtype}, we show periodic orbits for the second black hole solution. In this figure, $c_{13}=0$, and we set the value of $c_{14}$ to be 0, 0.00002, 0.2 from left to right in each column. We can see that in both figures the trajectory of the orbits shows exactly periodic behavior, which is determined by the given combinations of integers $(z,w,v)$. The integer $z$ governs the number of leaves in the orbital trajectory, $w$ determines the number of small loops within the orbit, and $v$ dictates the orientation (clockwise or counterclockwise) of the system.

\section{Gravitational Radiation in Einstein-\AE{}ther theory}\label{section4}
In this section, we provide a preliminary exploration
of the gravitational radiation emitted by the periodic orbits of a test particle orbiting supermassive Einstein-\AE{}ther black holes.

\subsection{Polarizations of gravitational waves in Einstein-\AE{}ther theory}

Let us first consider the timelike geodesic deviations to study the polarizations of gravitational waves in Einstein-\AE{}ther theory. In the spacetime described by the metric, $g_{\mu\nu}=\Lambda_{\mu\nu}+h_{\mu\nu}$, the spatial part, $\zeta_{i}$, takes the form~\cite{Gong:2018cgj}
\begin{equation}
    \ddot\zeta=-R_{0i0j}\zeta^{j}\equiv\frac{1}{2}\ddot{\mathcal{P}}_{ij}\zeta^{i},
\end{equation}
where $\zeta_{\mu}$ describes the deviation vector between two nearby trajectories of test particles and
\bqn
R_{0i0j} &\backsimeq& \frac{1}{2}(h_{0j,0i}+h_{0i,0j}-h_{ij,00}-h_{00,ij}) \nb\\
&=& \frac{1}{2}\left[ \vphantom{\frac{1}{2}} 2\dot\gamma_{(i,j)}+2\dot\gamma_{,ij}-2w^{0}_{,ij}-\ddot\phi_{ij} \right.\nb\\
&&\left.- 2\ddot\phi_{i,j}-\ddot\phi_{,ij}-\frac{1}{2}\delta_{i,j}\Delta \ddot{f}+\frac{1}{2}\ddot{f}_{,ij}\right] \nb\\
&=&-\frac{1}{2}\ddot{\phi}_{ij}+\dot{\Psi}^{\uppercase\expandafter{\romannumeral+2}}_{(ij)}+\Phi^{\uppercase\expandafter{\romannumeral+4}}_{,ij}-\frac{1}{2}\delta_{ij}\ddot{\Phi}^{\uppercase\expandafter{\romannumeral+2}},
\eqn
here $\Phi^{\uppercase\expandafter{\romannumeral+1}},\Phi^{\uppercase\expandafter{\romannumeral+2}},\Phi^{\uppercase\expandafter{\romannumeral+3}},\Phi^{\uppercase\expandafter{\romannumeral+4}}$ are four gauge-invariant quantities constructed from six scalar fields $(h_{00},w^0,\nu,\gamma,f,\phi)$, while $\Psi^{\uppercase\expandafter{\romannumeral+1}},\Psi^{\uppercase\expandafter{\romannumeral+2}}$ are gauge-invariant quantities constructed from three vector fields $(\gamma_i,\nu_i,\phi_i)$~\cite{Lin:2018ken}.

Assuming that $(\vec{e}_{X},\vec{e}_{Y},\vec{e}_{Z})$ are three unity vectors and form an orthogonal basis with $\vec{e}_{Z}\equiv \vec{N}$, so that $(\vec{e}_{X},\vec{e}_{Y})$ lies on the plane orthogonal to the propagation direction $\vec{N}$ of the gravitational wave, we find that in the coordinates $x^{\mu}=(t,x^{i})$, these three vectors can be specified by two angles, $ \theta $ and $ \phi$, via the relations.
\begin{align}
&\vec{e}_{X}=(\cos\theta \cos\phi,\cos\theta \sin\phi,-\sin\theta),\nb\\
&\vec{e}_{Y}=(-\sin\phi,\cos\phi,0),\nb\\
&\vec{e}_{Z}=(\sin\theta \cos\phi,\sin\theta \sin\phi,\cos\theta). 
\end{align}
Then we can define the six polarization $h_{N}$'s by
\begin{align}
    &h_{+}\equiv \frac{1}{2}(\mathcal{P}_{XX}-\mathcal{P}_{YY}),   h_{\times}\equiv \frac{1}{2}(\mathcal{P}_{XY}+\mathcal{P}_{YX}),\nb\\
&h_{b}\equiv \frac{1}{2}(\mathcal{P}_{XX}+\mathcal{P}_{YY}),   h_{L}\equiv \mathcal{P}_{ZZ},\nb\\
&h_{X}\equiv\frac{1}{2}(\mathcal{P}_{XZ}+\mathcal{P}_{ZX}),   h_{Y}\equiv \frac{1}{2}(\mathcal{P}_{YZ}+\mathcal{P}_{ZY}),
\end{align}
where $\mathcal{P}_{XY}\equiv \mathcal{P}_{ij}e^{i}_{X}e^{j}_{Y}$, and $\mathcal P_{ij}$ takes the form
\bqn
\mathcal{P}_{ij}&=&\phi_{ij}-\frac{2c_{13}}{(1-c_{13})c_V}\Psi^{\uppercase\expandafter{\romannumeral+1}}_i N_j \nb\\
&&-\frac{c_{14}-2c_{13}}{c_{14}(c_{13}-1)c^2_S}\Phi^{\uppercase\expandafter{\romannumeral+2}}N_iN_j+\delta_{ij}\Phi^{\uppercase\expandafter{\romannumeral+2}}
\eqn
In Einstein-\AE{}ther theory, all six polarization modes exist, but only five of them are independent: two from each of the vector and tensor modes, and one from the scalar mode.

The extra polarization modes in the Einstein-\AE{}ther theory introduce extra radiative channels which can significantly modify energy loss in the strong-field regime. In ref.~\cite{Foster:2007gr}, the strong-field radiation damping in binaries has been investigated in detail. That analysis motivates a detailed evaluation of the emitted waveform components in theories with an aether field. According to ref.~\cite{Zhang:2019iim, Lin:2018ken}, for a quasi-circular orbit, the emitted six polarization modes of gravitational waves can be expressed as
\bqn
\lb{h}
h_{+}&=&\frac{G_{æ}}{R}\ddot{Q}_{kl}e^{kl}_{+}, \nb\\
h_{\times}&=&\frac{G_{æ}}{R}\ddot{Q}_{kl}e^{kl}_{\times},\nb\\
h_{b}&=&\frac{c_{14}G_{æ}}{R(2-c_{14})}\left[3(Z-1)\ddot{Q}_{ij}e^{i}_{Z}e^{j}_{Z} \right. \nb\\
&&~~~~~~~~~~~~ \left.-\frac{4}{c_{14}c_{S}}\Sigma_{i}e^{i}_{Z}+Z\ddot{I}\right],\nb\\
h_{L}&=&\left[1-\frac{c_{14}-2c_{13}}{c_{14}(c_{13}-1)c^{2}_{S}}\right]h_{b},\nb\\
h_{X}&=&\frac{2c_{13}G_{æ}}{(2c_{1}-c_{13}c_{-})c_{V}R}[\frac{c_{13}\ddot{Q}_{jk}e^{k}_{Z}}{(1-c_{13})c_{V}}-2\Sigma_{j}]e^{j}_{X},\nb\\
h_{Y}&=&\frac{2c_{13}G_{æ}}{(2c_{1}-c_{13}c_{-})c_{V}R}[\frac{c_{13}\ddot{Q}_{jk}e^{k}_{Z}}{(1-c_{13})c_{V}}-2\Sigma_{j}]e^{j}_{Y},\nb\\
\eqn
where $R\equiv |\vec{x}|\gg d$, with $d$ denoting the size of the source. The dipolar quantity $\Sigma_i\equiv \int d^3xt_i$, with $t_i$ denotes the nonlinear source terms ref.~\cite{Foster:2006az}. $\ddot Q_{ij}$ is the second time derivative of the mass quadrupole moment and
\bqn
c^{2}_{S}&=&\frac{c_{123} (2-c_{14})}{c_{14}(1-c_{13})(2+c_{13}+3c_{2})},\\
c^{2}_{V}&=&\frac{2c_{1}-c_{13}(2c_{1}-c_{13})}{2c_{14}(1-c_{13})},\\
c^{2}_{T}&=&\frac{1}{1-c_{13}},
\eqn
and
\bqn
Z\equiv\frac{(\alpha_1-2\alpha_2)(1-c_{13})}{3(2c_{13}-c_{14})}.
\eqn
Here $\alpha_1$ and $\alpha_2$ are two post-Newtonian parameters \cite{Will:2014kxa}, which measure the preferred frame effects. Written in four dimensionless coupling constants, they are given by \cite{Foster:2005dk}
\begin{align}\lb{alpha}
\alpha_{1}&=-\frac{8(c_{1}c_{14}-c_{-}c_{13})}{2c_1-c_-c_{13}},\nb\\
\alpha_2&=\frac{1}{2}\alpha_1+\frac{(c_{14}-2c_{13})(3c_2+c_{13}+c_{14})}{c_{123}(2-c_{14})},   
\end{align}
where $c_{-}=c_{1}-c_{3}$ and
\begin{align}\lb{Sigma}
\Sigma\equiv \left(\alpha_1-\frac{2}{3}\alpha_2\right)\left(\frac{\Omega_1}{m_1}-\frac{\Omega_2}{m_2}\right),    
\end{align}
where $e^{ij}_+\equiv \vec{e}^i_X \vec{e}^j_X-\vec{e}^i_Y \vec{e}^j_Y$ and $e^{ij}_\times \equiv \vec{e}^i_X \vec{e}^j_Y+\vec{e}^i_Y \vec{e}^j_X$, and $\Omega_a$ denotes the binding energy of the $a$th compact body. 
From the above expressions, we can see that the scalar longitudinal mode $h_{L}$ is proportional to the scalar breather mode $h_{b}$. So, only five of the modes are independent. We emphasize that the waveform expressions derived above do not include the full 2.5 post-Newtonian (PN) radiation-reaction corrections to the waveform and to the orbital phasing. Recent work in ref.~\cite{Taherasghari:2025mlf} has computed radiation-reaction effects at 2.5PN order. Inclusion of these 2.5PN contributions is necessary for high-precision waveform modelling and for accurate long-term phase evolution (particularly for long inspirals), but is beyond the scope of the present weak-field/leading-order waveform calculation. We therefore leave the incorporation of the full 2.5PN corrections into our waveform and phasing model to future work.

As is mentioned, there are two static black hole solutions in Einstein-\AE{}ther theory, with $c_{14}=0$ and $c_{123}=0$ respectively. When $c_{14}=0$ (which corresponds to the first type Einstein-\AE{}ther black hole), all the extra polarization modes vanish identically and only the two plus and cross modes exist. In this case, we only need to consider the two tensorial modes $h_+$ and $h_\times$. When $c_{123}=0$, one can check that the two scalar polarization modes $h_L$ and $h_b$ apparently become divergent since $c_{123}$ appears in the denominators of the expressions of $h_L$ and $h_b$. This is because the propagation speeds of these two scalar modes become zero, which indicates that they are not propagating modes. In this case, the expressions of Eq.~(\ref{alpha}) for $h_b$ and $h_L$ are not applicable. To explore how the aether field affects the radiation of gravitational waves for the case when $c_{123}=0$, we restricted this study to the two tensorial modes $h_+$ and $h_\times$ and ignored all the other extra polarization modes. Observationally, all the GW signals observed by the LIGO/Virgo/KAGRA collaboration exhibit compatibility with the two tensor polarization modes, showing no statistically significant signatures of additional polarization modes \cite{KAGRA:2021vkt}. Even if additional modes exist, their effects are expected to be subdominant compared to the two tensorial modes. Based on the above reasons, in the next subsection, assuming that these additional modes are small, we focus exclusively on the radiation of the two traceless and transverse tensor modes of the gravitational waves for both types of Einstein-\AE{}ther black holes.

\subsection{GWs emitted from periodic orbits in Einstein-\AE{}ther theory}

In this section, we provide a preliminary exploration of the gravitational radiation emitted by the periodic orbits of a test particle orbiting a supermassive black hole in Einstein-\AE{}ther theory. Now we take into account an extreme-mass-ratio inspiral (EMRI) system, which is a system in which the smaller object has a mass much smaller than the supermassive black hole. The gravitational wave is generated when the object approaches the black hole, causing a loss of energy and angular momentum. Thus, the object gradually spirals and falls into the black hole. In such systems, the energy and angular momentum losses caused by the motion of the smaller mass object are negligible compared to the total energy over a few orbital periods, enabling the validity of the adiabatic approximation. The adiabatic approximation is valid when the inspiral timescale is much longer than the orbital period, so that the object's energy and angular momentum can be reckoned as constant over several orbits, allowing us to neglect the effect of gravitational radiation. Under this approximation, the trajectory of the smaller object can be described by geodesic equations, and the resulting GW waveforms can be derived using the formulas in Eqs.~(\ref{h}) for $h_+$ and $h_\times$. 

The main procedure of adiabatic approximation is introduced as follows. Firstly, we treat the smaller object as a test particle, and calculate the periodic orbits of the particle by solving the geodesic equation. Secondly, we use the formula of gravitational radiation to obtain the corresponding gravitational waves. This approach allows us to explore crucial information about the fundamental properties and physics of both the orbits and the central black hole, as well as the \ae{}ther effect. 

The waveforms given in (\ref{h}) are derived from the trajectory of the object in the Schwarzschild coordinate system $(r, \theta, \phi)$, but the waveform is typically expressed in a detector-adapted coordinate system $(X, Y, Z)$. The transformation from Schwarzschild coordinates to Cartesian coordinates is given by~\cite{Babak:2006uv}
\bqn
x=r \sin\theta \cos\phi,\ y=r \sin\theta \sin\phi, \ z=r\cos\theta.
\eqn
For a small object with mass $m$ following a trajectory $Z^i(t)$, the $Q_{ij}$ is given by~\cite{Thorne:1980ru}
\bqn\lb{Q}
Q^{ij}=m\int x^ix^j\delta^3(x^i-Z^i(t))d^3x.
\eqn
To calculate the waveform, we recall the formulas in Eqs.~(\ref{h}) for the plus $h_+$ and cross $h_\times$ polarizations of the GW,
\bqn
h_{+}&=&\frac{G_{æ}}{D_{\rm L}}\ddot{Q}_{ij}e^{ij}_{+}, \\
h_{\times}&=&\frac{G_{æ}}{D_{\rm L}}\ddot{Q}_{ij}e^{ij}_{\times},
\eqn
with $D_{\rm L}$ being the luminosity distance. 

To illustrate the GW waveforms of different periodic orbits and how the \ae{}ther effect can affect it, we consider an EMRI system consisting of a small object with mass $m=10 M_\odot$ and a supermassive black hole with mass $M=10^{7} M_\odot$, where $M_\odot$ is the solar mass. For simplicity, we set the inclination angle $\Theta=\pi/4$ and the latitude angle $\Phi=\pi/4$. The luminosity distance is set to $D_{\rm L}=200\;{\rm Mpc}$.

\begin{figure*}[htbp]
    \centering
    \begin{minipage}{0.49\linewidth}
		\centering
		\includegraphics[width=0.75\linewidth]{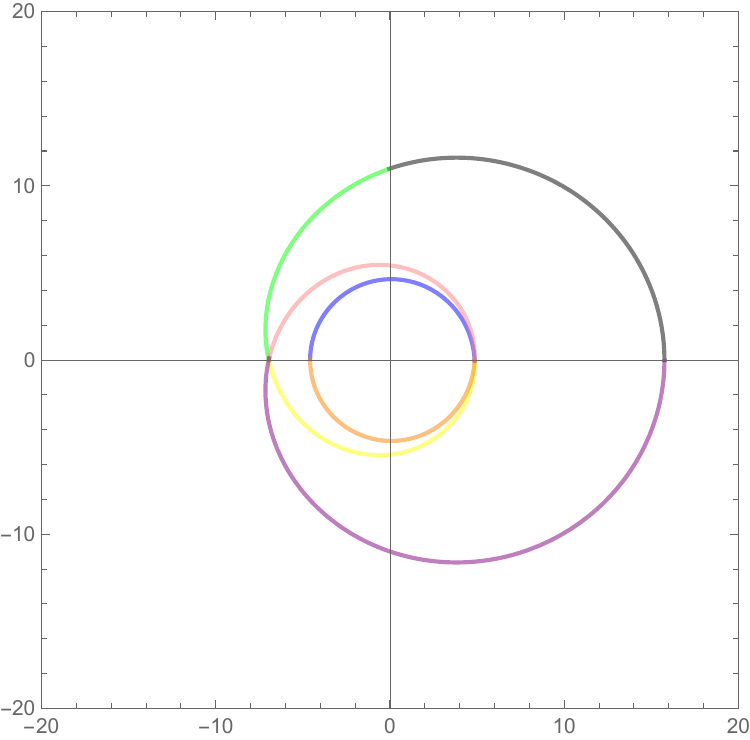}\hspace{15mm}
		\label{orbit120}
	\end{minipage}
    \begin{minipage}{0.49\linewidth}\vspace{8mm}\hspace{5mm}
		\centering
		\includegraphics[width=9cm,height=6cm]{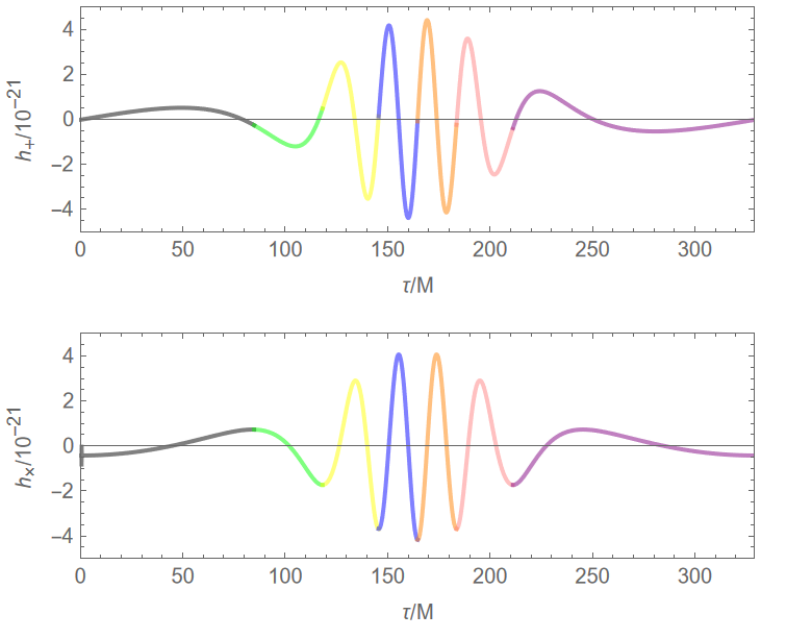}
		\label{hcross}
	\end{minipage}
    \caption{The left figure is a figure showing a particle traveling from an apastron to another in a typical periodic orbit around a second type of Einstein-\AE{}ther black hole with $(z, w, v) = (1, 2, 0)$. The right figure represents the $h_{+}$ and $h_{\times}$ mode of gravitational waves of a black hole in Einstein-aether theory, with $q=2$. In both figures, different colors represent each period of their orbits. In the figure of the left panel, the horizontal and vertical axes are Cartesian coordinates $(x, y) =(r\sin\phi, r \cos\phi)$ with $\theta=\pi/2$.}
    \label{120waveform}
\end{figure*}

\begin{figure*}[htbp]
    \centering
    \begin{minipage}{0.49\linewidth}
		\centering
		\includegraphics[width=0.75\linewidth]{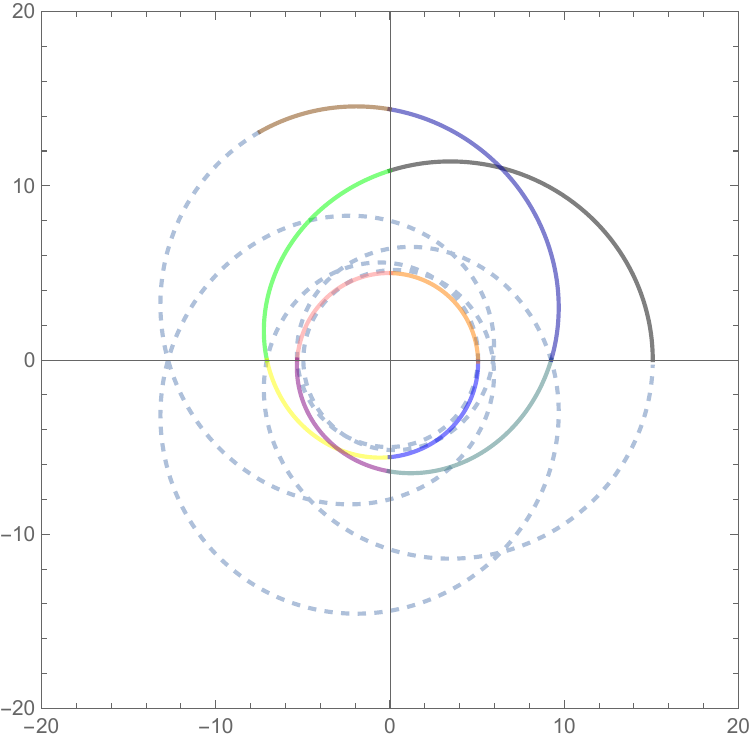}\hspace{15mm}
		\label{orbit311}
	\end{minipage}
    \begin{minipage}{0.49\linewidth}\vspace{4mm}\hspace{5mm}
		\centering
        \includegraphics[width=9cm,height=6cm]{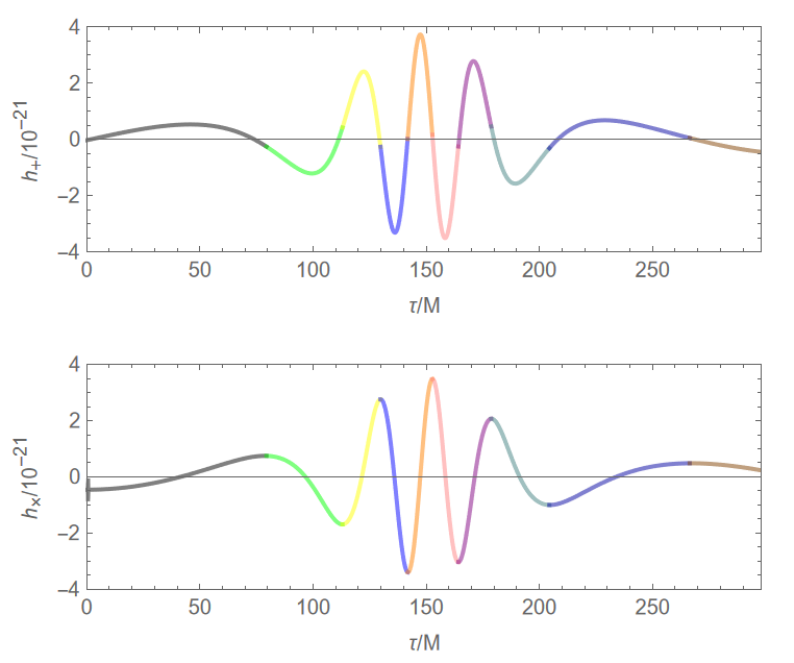}
	\end{minipage}
    \caption{The left figure is a figure showing a particle traveling from an apastron to another in a typical periodic orbit around a second type of Einstein-\AE{}ther black hole with $(z, w, v) = (3, 1, 1)$. The right figure represents the $h_{+}$ and $h_{\times}$ mode of gravitational waves of a black hole in Einstein-aether theory, with $q=\frac{4}{3}$. In both figures, different colors represent each period of their orbits. In the figure of the left panel, the horizontal and vertical axes are Cartesian coordinates $(x, y) =(r\sin\phi, r \cos\phi)$ with $\theta=\pi/2$.}
    \label{311waveform}
\end{figure*}

In Figs.~\ref{120waveform} and~\ref{311waveform}, we show waveforms emitted by two typical periodic orbits in the second type of Einstein-\AE{}ther black hole, with $(z,w,v) = (1,2,0)$ and $(z,w,v) = (3,1,1)$. In both figures, we utilize various colors to emphasize the connection between the gravitational waveforms and the periodic orbit, facilitating a comprehensive examination of how the periodic orbit influences the resulting gravitational radiation. We can see that the GW waveform shows zoom-whirl behaviors in an entire circle. By comparing the figures of the periodic orbits with the figures of the GW waveforms, we can see that the amplitude of the GW form increases as the object travels into the perihelion and decreases as it moves farther from the perihelion. In the beginning stage of the trajectory, the gravitational field is weak as the small object is far from the black hole, thus the GW waveforms are smooth. When the small object travels into the strong field regime, the trajectory gets distorted, with an evident change in both the frequency and amplitude of the GW. In the final stage, the small object is near the event horizon, and the GW signal reaches its peak with a sharp burst in both frequency and amplitude. Consequently, the periodic orbits and their corresponding GW signals carry significant information about the EMRI system, which can be helpful in understanding the physical properties of black holes and their surrounding areas. 

\begin{figure*}[htbp]
    \centering
    \begin{minipage}{0.49\linewidth}
		\centering
		\includegraphics[width=0.75\linewidth]{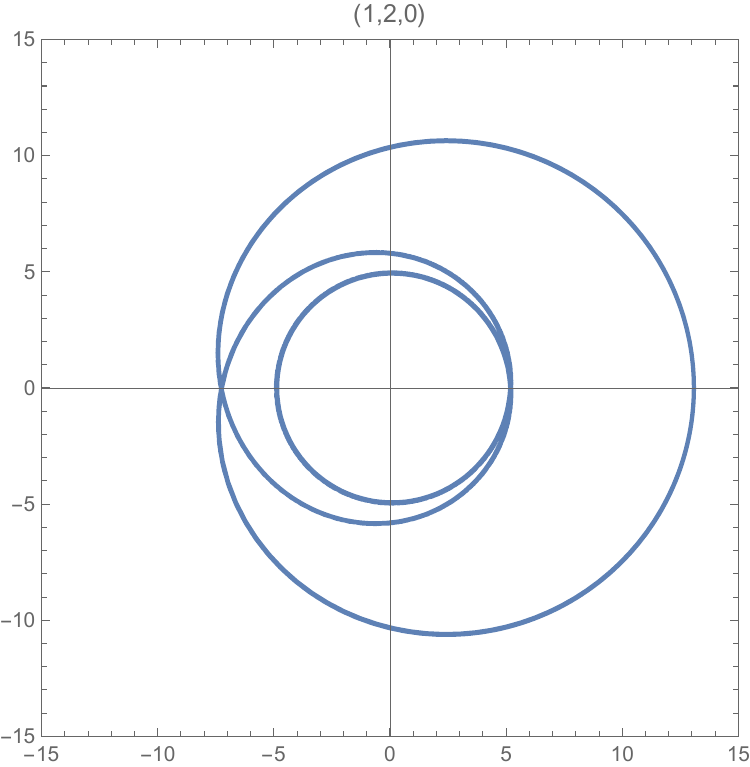}\hspace{15mm}
	\end{minipage}
    \begin{minipage}{0.49\linewidth}
		\centering
        \includegraphics[width=9cm,height=6cm]{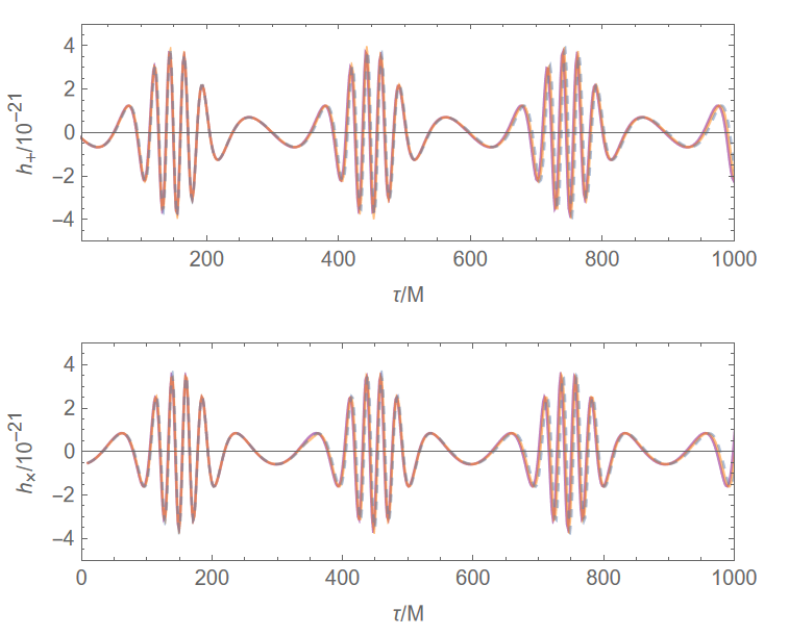}\vspace{-8mm}\hspace{-5mm}
	\end{minipage}
    \caption{The left figure is a figure showing a typical periodic orbit around a first type of  Einstein-\AE{}ther black hole with $(z, w, v) = (1, 2, 0)$. The right figure represents the $h_{+}$ and $h_{\times}$ modes of gravitational waves of a black hole in Einstein-\AE{}ther theory, with $q=2$. The dashed line represents the GW of the Schwarzschild black hole, namely $c_{13}=0$. The orange line represents the GW of Einstein-\AE{}ther black hole with $c_{13}=0.05$. The purple line represents the GW of Einstein-\AE{}ther black hole with $c_{13}=0.1$. In the figure of the left panel, the horizontal and vertical axes are Cartesian coordinates $(x, y) =(r\sin\phi, r \cos\phi)$ with $\theta=\pi/2$.}
    \label{120GWdifferencec13}
\end{figure*}

\begin{figure*}[htbp]
    \centering
    \begin{minipage}{0.49\linewidth}
		\centering
		\includegraphics[width=0.75\linewidth]{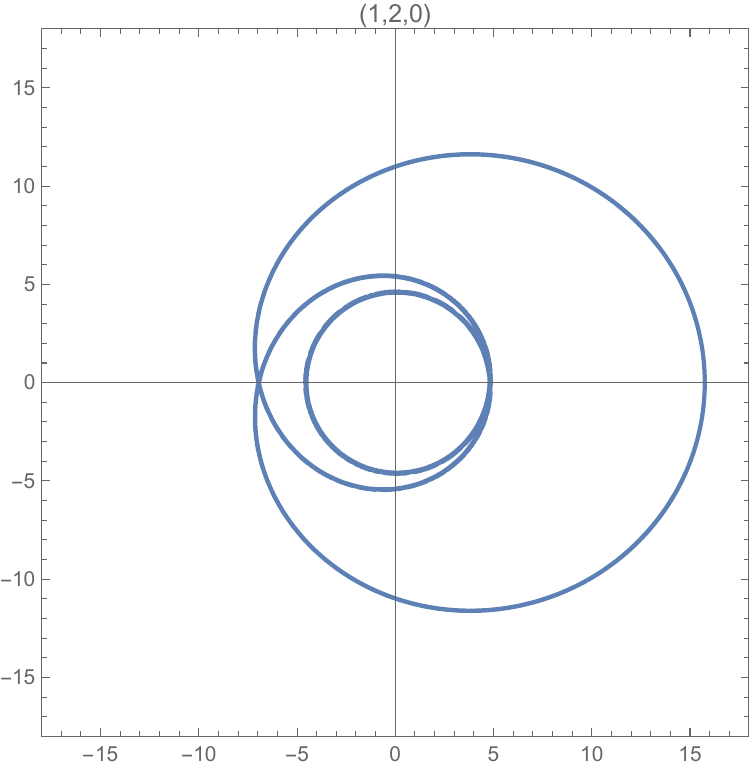}\hspace{15mm}
	\end{minipage}
    \begin{minipage}{0.49\linewidth}
        \centering
        \includegraphics[width=9cm,height=6cm]{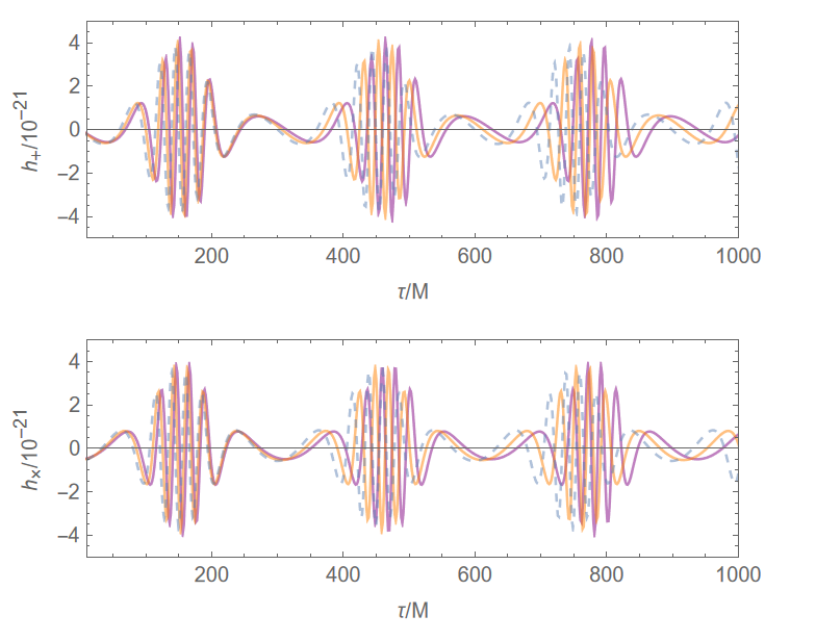}\vspace{-8mm}\hspace{-5mm}
	\end{minipage}
    \caption{The left figure is a figure showing a typical periodic orbit around a second type of Einstein-\AE{}ther black hole with $(z, w, v) = (1, 2, 0)$. The right figure represents the $h_{+}$ and $h_{\times}$ modes of gravitational waves of a black hole in Einstein-\AE{}ther theory, with $q=2$. The dashed line represents the GW of the Schwarzschild black hole, namely $c_{14}=0$. The orange line represents the GW of Einstein-\AE{}ther black hole with $c_{14}=0.05$. The purple line represents the GW of Einstein-\AE{}ther black hole with $c_{14}=0.1$. In the figure of the left panel, the horizontal and vertical axes are Cartesian coordinates $(x, y) =(r\sin\phi, r \cos\phi)$ with $\theta=\pi/2$.}
    \label{120GWdifference}
\end{figure*}

In Figs.~\ref{120GWdifferencec13} and~\ref{120GWdifference}, we plot typical periodic orbits with $(z,w,v)=(1,2,0)$ and their $h_+$ and $h_\times$ modes with different $c_{14}$ and $c_{13}$, corresponding to two different Einstein-\AE{}ther black holes. However, it is still worth mentioning that Fig.~\ref{120GWdifferencec13} is plotted only for theoretical illustration, as the multi-messenger event GW170817/GRB 170817A gives the constraint of $c_{13}\leq10^{-15}$ \cite{Oost:2018tcv, Gong:2018cgj}. First, a clear relationship emerges between waveform and orbit in which the quiet phases correspond to ``leaves", while intense oscillations align with whirl-dominated segments of the trajectory. Secondly, the \ae{}ther effect of $c_{13}$ is relatively weak in Fig.~\ref{120GWdifferencec13}, as the changes in amplitude and phase are negligible. In Figs.~\ref{120GWdifference}, compared to the Schwarzschild case with $c_{14}=0$, the \ae{}ther parameter $c_{14}$ mainly affects the phase of the GW signals and gives a slight change in amplitudes. As $c_{14}$ increases, the amplitudes increase slightly. These findings offer potential for detecting and distinguishing physical signatures in future detections.

\section{Discussions and Conclusions}\label{section5}

In this paper, we study the periodic orbits and corresponding waveforms in Einstein-\AE{}ther theory. From the geodesic equations for two different types of black holes, we can analytically solve the equation for the first type. Then we use a special taxonomy \cite{Levin:2008mq} to distinguish different types of periodic orbits in Einstein-\AE{}ther theory. In this scheme, each periodic orbit is described by a set of parameters $(z,w,v)$. In Figs.~\ref{firsttype} and \ref{secondtype}, we plot periodic orbits in two types of black holes in Einstein-\AE{}ther theory with different \ae{}ther parameter $(c_{13},c_{14})$. 

The radiation of gravitational waves from periodic orbits in Einstein-\AE{}ther theory is preliminarily considered. Einstein-\AE{}ther theory is known for its five independent polarizations, although only $h_{+}$ and $h_{\times}$ remain possible for the two static spherical solutions. Figs.~\ref{120waveform} and ~\ref{311waveform} use different colors to highlight the correlation of periodic orbits and their GW waveforms, enabling us to have a better understanding of the relationship between periodic orbits and the resulting gravitational waves. We can see that in both figures the GW waveform amplitude peaks when the object is near the ``whirl'' part of the orbit and decreases as the object moves to the ``zoom'' part of the orbit. During the initial phase, when the object resides at large distances from the black hole within the weak-field regime, the waveform characteristics manifest as a smooth, low-amplitude signal. As the object enters the strong field region near the black hole, its trajectory gets distorted, and both the frequency and the amplitude of the waveform increase sharply. In the final phase of motion, as the object travels near the horizon, the GW signal reaches its peak. As discussed above, GWs obviously carry crucial information about the particle trajectory, mass distribution, and fundamental properties, which can be helpful in understanding the physical characteristics of black holes. It is also shown in Figs.~\ref{120GWdifferencec13} and~\ref{120GWdifference} that \ae{} the parameters $c_{13}$ and $c_{14}$ mainly affect the phases of GW rather than their amplitudes. It is also shown that the \ae{}ther effect is stronger in the second solution than in the first one. These results may provide us with a way to differentiate between the two types of black holes in Einstein-\AE{}ther theory and the Schwarzschild black hole.

\section*{Acknowledgements}

We appreciate the helpful discussions with Dr. Xiao Zhi, Bo-Yang Zhang, and Yuan-Zhu Wang. 
This work is supported by the National Key Research and Development Program of China under Grant No. 2020YFC2201503, the National Natural Science Foundation of China under Grants No. 12275238, and the Zhejiang Provincial Natural Science Foundation of China under Grants No. LR21A050001 and No. LY20A050002.


\appendix


%

\end{document}